\documentclass[jmp,twocolumn,preprintnumbers,aps,superscriptaddress,amsmath,amssymb]{revtex4}

\usepackage{graphicx}
\usepackage{xspace}
\usepackage{dcolumn}
\usepackage{bm}
\usepackage{amscd}
\usepackage{amsthm}
\usepackage{german}
\usepackage[T1]{fontenc}
\usepackage{color}
\usepackage{pstricks}

\input xy
\xyoption{all}

\newcommand{\unity}{\ensuremath{{\rm 1 \negthickspace l}{}}}

\renewcommand{\vec}{\operatorname{vec}}
\newcommand{\tr}{\operatorname{tr}}

\newcommand{\reach}{\operatorname{Reach}}

\newcommand{\ad}[1]{\operatorname{ad}_{#1}}
\newcommand{\Ad}[1]{\operatorname{Ad}_{#1}}

\newcommand{\C}{\mathbb{C}}

\newcommand{\N}{\mathbb{N}}

\newcommand{\bG}{\mathbf{G}}

\newcommand{\bK}{\mathbf{K}}

\newcommand{\bS}{\mathbf{S}}
\newcommand{\bP}{\mathbf{P}}
\newcommand{\bT}{\mathbf{T}}

\newcommand{\SU}{\mathrm{SU}}

\newcommand{\bch}{{\sc{bch}}\xspace}

\newcommand{\adr}{\operatorname{ad}}
\newcommand{\Adr}{\operatorname{Ad}}

\newcommand{\su}{\mathfrak{su}}

\newcommand{\gl}{\mathfrak{gl}}

\newcommand{\e}{{\rm e}}
\newcommand{\ri}{{\rm i}}

\newcommand{\rL}{{\rm L}}
\newcommand{\rE}{{\rm E}}
\newcommand{\g}{\mathfrak{g}}

\newcommand{\R}{\mathbb{R}}

\newcommand{\expt}[1]{\ensuremath{\langle #1 \rangle}{}}

\newcommand{\argmax}{\operatorname{argmax}{}}

\theoremstyle{plain}

\newtheorem{corollary}{Corollary}[section]
\newtheorem{lemma}{Lemma}[section]
\newtheorem{proposition}{Proposition}[section]
\newtheorem{theorem}{Theorem}[section]

\theoremstyle{definition}

\newtheorem{remark}{Remark}[section]

\theoremstyle{remark}

\begin{document}


\title{Lie-Semigroup Structures for Reachability and Control of Open Quantum Systems:\\
{{Kossakowski-Lindblad Generators Form Lie Wedge to Markovian Channels}}}

\author{G.~Dirr}\email{dirr@mathematik.uni-wuerzburg.de}
\affiliation{Institute of Mathematics, W{\"u}rzburg University,
Am Hubland, D-97074 W{\"u}rzburg, Germany}

\author{U.~Helmke}
\affiliation{Institute of Mathematics, W{\"u}rzburg University,
Am Hubland, D-97074 W{\"u}rzburg, Germany}

\author{I.~Kurniawan}
\affiliation{Institute of Mathematics, W{\"u}rzburg University,
Am Hubland, D-97074 W{\"u}rzburg, Germany}

\author{T.~Schulte-Herbr{\"u}ggen}\email{tosh@ch.tum.de}
\affiliation{Department of Chemistry, Technical University Munich,
Lichtenbergstrasse 4, D-85747 Garching, Germany}

\date{\today}

\begin{abstract}
\noindent
In view of controlling finite dimensional open quantum systems,
we provide a unified Lie-semigroup framework describing
the structure of completely positive trace-preserving maps.
It allows (i)~to identify the Kossakowski-Lindblad generators as
the Lie wedge of a subsemigroup, (ii) to link 
properties of Lie semigroups such as divisibility with Markov
properties of quantum channels,
and (iii) to characterise reachable sets and controllability in open systems.
We elucidate when time-optimal
controls derived for the analogous closed system  already give good fidelities
in open systems and when a more detailed knowledge of the open system (e.g., in terms of the
parameters of its Kossakowski-Lindblad master equation)
is actually required for state-of-the-art optimal-control algorithms.
As an outlook, we sketch the structure of a new, potentially more efficient
numerical approach explicitly making use of the corresponding Lie wedge.\\[3mm]


\end{abstract}
\maketitle


\section{Introduction}
\label{Sec:intro}

 Understanding and manipulating open quantum systems and quantum
channels is an important challenge for exploiting quantum effects in
future technology \cite{DowMil03}.

Protecting quantum systems against relaxation 
is therefore tantamount to using coherent superpositions as a resource.
To this end, decoherence-free subspaces have been applied \cite{ZR97LCW98},
bang-bang controls \cite{VKLabc} have been used for decoupling the
system from dissipative interaction with the environment, while a
quantum Zeno approach \cite{MisSud77} may be taken to projectively
keep the system within the desired subspace \cite{FacPas02}. 
Very recently, the opposite approach has been taken by
solely expoiting relaxative processes for state 
preparation \cite{Wolf08c,Zoller08a}.
It is an extreme case of engineering quantum dynamics in open systems
\cite{VioLloyd01}, where targeting fix points has lately
become of interest \cite{BNT08ab}.

In either case, for exploiting the power of
system and control theory, first the quantum
systems has to be characterised, e.g., by input-output relations in the
sense of quantum process tomography. Deciding whether the dynamics of
the quantum system thus specified allows for a Markovian description
to good approximation (maybe up to a certain level of noise) has recently
been addressed \cite{Wolf08a,Wolf08b,Wolf08pc}. 
This is of crucial interest,
since a Markovian equation of motion paves the way to applying
the power Lie-theoretic methods \cite{JS72,Jurdjevic97}
from geometric and bilinear control theory. 
Moreover, it comes with the
well-established frameworks of completely positive semigroups
and Kraus representations \cite{Kraus71, Koss72, Koss72b, Choi75, GKS76,
Lind76, Kraus83,Rabitz07b}.

On the other hand, the specific {\em Lie-semigroup} aspects of open quantum systems
clearly have not
been elaborated on in the pioneering period 1971--76 of completely positive 
semigroups \cite{Koss72, Koss72b, GKS76, Lind76, Davies76}, 
mainly since major progress in the understanding of
Lie semigroups was made in the 
decade 1989--99 \cite{HHL89, Egg-Diss, Neeb_LieSG92, LNM1552, EM20, HofRupp97div}.
While relations of Lie semigroups and classical control theory
were soon established,
e.g., in \cite{JurdKupka81a, JurdKupka81b, LNM998, HofRupp91int, Mitt-Diss, EM20, Mitt95, Lawson99}, 
only recently the use of Lie-semigroup terms in the {\em control of open quantum systems} 
was initiated \cite{Alt03,DiHeGAMM08}, where in \cite{Alt03} the elaborations
were confined to single two-level systems.
However, we see a great potential in exploiting the algebraic
structure of Lie-semigroup theory for practical problems of
reachability and control of open quantum systems.

Its importance becomes evident, because
among the generic tools needed for the current advances in quantum
technology (for a survey see, e.g., \cite{DowMil03}), quantum control
plays a major role. From formal description of quantum optimal control \cite{Sam90+} the theoretical
aspects of existence of optima soon matured into numerical algorithms
solving practical problems of steering quantum dynamics \cite{Rabitz87,Rabitz90,Krotov,GRAPE}.
Their key concern is to find optima of some quality
function like the quantum gate fidelity under realistic conditions
and, moreover, constructive ways of achieving those optima given the
constraints of an accessible experimental setting. For a recent introduction,
see~\cite{dAless08}. However, realistic
implementations in open quantum systems are mostly beyond analytical
tractability. Hence numerical methods are often indispensible,
where gradient-like algorithms are the most basic, but robust tools.
Thus they proved applicable to a broad array of problems including
optimal control of closed quantum systems \cite{GRAPE,NMRJOGO06}
and computing entanglement measures \cite{SGDH08,GAMM06,CuDiHeICIAM07}.
For mathematical details on gradient systems as numerical tools for
constrained optimisation, we refer to  \cite{Bro88+91,Bloch94, HM94}.

Generalising these well-established gradient techniques,
in our previous work \cite{SGDH08}, we have exploited the geometry of
Riemannian manifolds related to Lie groups, their subgroups, and
homogeneous spaces in a common framework for setting up gradient
flows in closed quantum systems. 
There we addressed (a) {\em abstract optimisation tasks} on smooth
state-space manifolds and (b) {\em dynamic optimal control tasks} in the
specific time scales of an experimental setting. 
Here, we will see that the corresponding abstract optimisation tasks
for open quantum systems are much more involved, while the dynamic optimal
control tasks remain in principle the same. From a mathematical point
of view, this difficulty results from the fact that the evolution
of a controlled open quantum system is no longer described by
a semigroup of unitary propagators, i.e. by a semigroup 
contained in a \emph{compact} Lie group.


Thus, we extend the Lie-theoretic approach in \cite{SGDH08} to
finite dimensional {\em open quantum systems} and discuss their dynamics
in terms of Lie {\em semigroups}.
In particular, we characterise the Lie properties (the Lie wedge) of Markovian quantum
channels from the viewpoint of divisibility and local divisibility in semigroups. ---
On a general scale and with regard to practical applications of quantum control, 
knowing about the Lie-semigroup structure of the dynamic system
is shown to be highly advantageous: analysing its
tangent cones (Lie wedges) allows for addressing problems of
reachability, accessibility, controllability and actual control 
in a unified frame providing powerful Lie algebraic terms.


\section*{Starting Point}

To begin with, we briefly indicate how the theory elucidated in
previous work \cite{SGDH08} can be extended to reachable sets of non
necessarily controllable systems. In particular, we concentrate on
the structure of reachable sets and obstacles arising from it.
Moreover, pertinent applications to open relaxative quantum dynamical systems 
are elaborated---proving the relevance of the semigroup setting in physics.

The starting point in \cite{SGDH08} was a smooth state-space manifold
$M$ or a controllable dynamical system on $M$, i.e.~a control system
whose \emph{reachable sets} $\reach(X_0)$ satisfy $\reach(X_0) = M$
for all $X_0 \in M$. For a right invariant system \eqref{invControlSystem}
the state space of which is given by a connected Lie group
$\mathbf G$, controllability is
equivalent to the fact that the entire group $\mathbf G$ can be
reached from the unity $\unity$, i.e.
\begin{equation}
\label{reachabilityI}
\bG = \reach(\unity) := 
\bigcup_{T \geq 0} \reach(\unity, T),
\end{equation} 
where $\reach(\unity, T)$ denotes the reachability set
in time $T \geq 0$, i.e.~the set of all states to where the systems can be
steered from $\unity \in \mathbf G$ in time $T$, cf. Eqn.\eqref{reachset}.
In general, however, we cannot expect Eqn.(\ref{reachabilityI}) to hold.
Nevertheless, the reachability sets $\reach(\unity, T_1)$ and
$\reach(\unity, T_2)$ of right invariant systems
obey the following multiplicative structure
\begin{equation*}
\reach(\unity, T_1)\cdot\reach(\unity, T_2)
= \reach(\unity, T_1+T_2).
\end{equation*} 
Thus $\reach(\unity)$ is a subsemigroup of $\bG$, see 
Sec.\ref{SSec:Reach}. --- Now, we will give a basic survey 
on subsemigroups and some of their applications in quantum control.


\section{Fundamentals of Lie Subsemigroups and Reachable Sets}
\label{sec:basics}


\subsection{Lie Subsemigroups}

For the following basic definitions and results on Lie
subsemigroups we refer to \cite{LNM1552, HHL89, Hof91sem-I, Egg91sem-II, Nee91sem-III}. However, the
reader should be aware of the fact that the terminology in this
area is sometimes inconsistent. Here, we primarily adopt the
notions used in \cite{LNM1552}. For further reading we also
recommend \cite{Lawson99}. 

A {\em subsemigroup} of a (matrix) Lie group $\bG$ with Lie algebra
$\g$ is a subset $\bS \subset \bG$ which contains the unity $\unity$
and is closed under multiplication, i.e. 
$\bS\cdot \bS \subseteq \bS$. The largest subgroup contained in
$\bS$ is denoted by $\rE(\bS) := \bS \cap \bS^{-1}$. The
\emph{tangent cone} of $\bS$ is defined as
\begin{equation*}
\rL(\bS) := 
\{ \dot{\gamma}(0) \;|\; 
\mbox{$\gamma(0) = \unity$, $\gamma(t) \in \bS$, $t\geq 0$}\}
\subset \g,
\end{equation*}
where $\gamma: [0,\infty) \to \bG$ denotes any smooth curve contained
in $\bS$. In order to relate subsemigroups to their tangent cones,
we need some further terminology from convex analysis.
A closed convex cone $\frak{w}$ of a finite dimensional real vector space
is called a \emph{wedge}. 

Moreover, a wedge $\frak{w}$ in a Lie algebra $\frak{g}$ is termed a 
{\em Lie semialgebra} if the wedge $\frak{w}$ is locally compatible
with the Baker-Campbell-Hausdorff (\bch) multiplication
$X*Y := X + Y + \tfrac{1}{2}[X,Y]+\dots$, defined
via the \bch series.
More precisely, there has to be an open \bch neighbourhood
$B \subset \frak{g}$ of $0$ such that $\frak{w}$ is locally invariant
under $*$, i.e.
\begin{equation}\label{eqn:Lie-semialg}
(\frak w \cap B) * (\frak w \cap B) \subseteq \frak w.
\end{equation}
For a thorough treatment of the \bch multiplication and Lie
semialgebras see~\cite{HHL89,Egg-Diss}.

The \emph{edge} of $\frak{w}$ denoted by
$\rE(\frak{w})$ is the largest subspace contained in $\frak{w}$,
i.e.~one has $\rE(\frak{w}) := \frak{w} \cap (-\frak{w})$.
Finally, a wedge $\frak{w}$ of a finite dimensional real (matrix)
Lie algebra $\frak{g}$ is called a \emph{Lie wedge} if it is
invariant under the group of inner automorphisms
$\mathbf{Inn(\frak{w})} := \langle \exp(\ad{\rE(\frak{w})}) \rangle$.
More precisely,
\begin{equation*}
\e^{\ad{g}}(\frak{w}) := \e^g\, \frak{w}\, \e^{-g} = \frak{w}
\end{equation*}
for all $g \in \rE(\frak{w})$.  Here and in the sequel, we denote by
$\langle M \rangle$ and  $\langle M \rangle_S$ the group and,
respectively, semigroup generated by the subset $M \subset \bG$.

\begin{remark}\label{rem:LieW-LieSA}
While every Lie semialgebra is also a Lie wedge, the converse does
in general not hold, as will be of importance in the paragraph on
divisibility in  Sec.~\ref{sec:LieS-Div}.
\end{remark}

Now, the fundamental properties
of the tangent cone $\rL(\bS)$ can be summarised as follows.

\begin{lemma}
\label{lemLiewedge}
Let $\bS$ be a closed subsemigroup of a Lie group
$\bG$ with Lie algebra $\g$ and let  $\frak{w} \subset \g$ be
any Lie wedge. Then the following statements are satisfied.
\begin{enumerate}
\item[(a)]
The edge of $\frak{w}$, $\rE(\frak{w})$, carries the structure of
a Lie subalgebra of $\g$.
\item[(b)]
The tangent cone $\rL(\bS)$ coincides with
\begin{equation}
\label{Liewedge2}
\rL(\bS) = \{ g \in {\mathfrak g}\,|\,
\mbox{$\exp (t g) \in \bS$ for all $t\geq 0$}\}.
\end{equation}
In particular, $\rL(\bS)$ is a Lie wedge of $\g$
which is $\mathrm{Ad}_{\rE(\bS)}$-invariant, i.e. 
$G \frak{w} G^{-1} = \frak{w}$
for all $G \in \rE(\bS)$. 
\item[(c)]
The edge of $\rL(\bS)$ fulfills the equality
$\rE(\rL(\bS)) = \rL(\rE(\bS))$.
\end{enumerate}
\end{lemma}

{\bf Proof.}\hspace{.5em}
\begin{enumerate}
\item[(a)]
Note that $\e^{t\ad{g}}(h) \in \rE(\frak{w})$ for all $t \in \R$
and $g,h \in \rE(\frak{w})$. Hence
\begin{equation*}
\frac{\rm d}{{\rm d}t}\e^{t \ad{g}} (h)\big|_{t=0}
= \ad{g}h \in \rE(\frak{w})
\end{equation*}
for all $g,h \in \rE(\frak{w})$, thus $\rE(\frak{w})$ is a Lie
subalgebra.
\item[(b)]
The proof of Eqn.~\eqref{Liewedge2} is rather technical and
therefore we refer to \cite{HHL89}, Proposition IV.1.21.
Once Eqn.~\eqref{Liewedge2} is established, one has
\begin{equation*}
\rL(\bS) = \bigcap_{t>0} t^{-1} \exp^{-1}(\bS)
\end{equation*}
and thus the continuity of the exponential map implies that
$\rL(\bS)$ is closed.
To see that $\rL(\bS)$ is a wedge we have to show:
(i) $\mu \rL(\bS) = \rL(\bS)$ for all $\mu \in \R^+$
and 
(ii) $\rL(\bS) + \rL(\bS) \subset \rL(\bS)$.
Property (i) is obvious; property (ii) follows by the
Trotter product formula
\begin{equation*}
\e^{t(g+h)} = 
\lim_{n\to\infty}\big(\e^{tg/n}\e^{th/n}\big)^n.
\end{equation*}
Finally, let $g \in \rE(\rL(\bS))$ and $h \in \rL(\bS)$, then
\begin{equation*}
\e^{g} e^{t h} \e^{-g}
= \exp\big(t\,\e^{g} \,h\, \e^{-g}\big) \in \bS
\end{equation*}
for all $t \geq 0$. Thus
$\e^{g} h \e^{-g} = \e^{\ad{g}}(h) \in \rL(\bS)$. The same argument
applies to $G \in \rE(\bS)$.
\item[(c)]
Let $g \in \rE(\rL(\bS))$. Then $\e^{tg} \in \bS$ for all $t\in\R$.
Thus $\e^{tg} \in \rE(\bS)$ and hence $g \in \rL(\rE(\bS))$. Therefore,
we have shown $\rE(\rL(\bS))\subset \rL(\rE(\bS))$.
The converse, $\rL(\rE(\bS)) \subset \rE(\rL(\bS))$, holds by definition.

{\hspace{-6mm}}For more details, see Proposition 1.14 in \cite{LNM1552}.
\hfill $\blacksquare$
\end{enumerate}

\medskip

For closed subsemigroups, Lemma \ref{lemLiewedge} provides the
justification to call the tangent cone $\rL(\bS)$ 
\emph{Lie-} or \emph{Lie-Loewner wedge} of $\bS$. 

\medskip

Unfortunately, the `\/local-global-correspondence\/' between Lie wedges
and (closed) connected subsemigroups is not as simple as the
correspondence between Lie subalgebras and Lie subgroups. On the one
hand, there are Lie wedges $\frak{w}$ such that `\/the\/' corresponding
subsemigroup $\bS$ is not unique, i.e.~the equality
$\frak{w} = \rL(\bS)$ holds for more than one subsemigroup $\bS$.
On the other hand, there are Lie wedges $\frak{w}$ which do not
act as Lie wedge of any subsemigroup, i.e.~$\frak{w} = \rL(\bS)$
fails for each subsemigroup $\bS$, cf. \cite{LNM1552}. 

Another subtlety in the theory of semigroups arises from the fact
that there may exist elements in $\bS$ that are arbitrarily close to the
unity but do {\em not} belong to any one-parameter semigroup 
completely contained in $\bS$ (a standard example being
a certain subsemigroup of the  Heisenberg group \cite{HHL89,HofRupp97div}).
This somewhat striking feature arises whenever the \bch multiplication
leads outside the Lie wedge $\rL(\bS)$. It does not occur as
soon as $\rL(\bS)$ also carries the structure of a Lie semialgebra,
cf.~Theorem~\ref{thm:LieSA-locdiv} below.
In general, however, the exponential map of a zero-neighbourhood in $\rL(\bS)$
need {\em not} give a $\unity$"~neighbourhood
in the semigroup.

Meanwhile, the following terminology is well-established
\cite{DokHof97,HofRupp97div}: a set $E \in \bG$ is called \emph{exponential}
if to each element $T \in E$ there exists a Lie algebra element
$g \in \mathfrak{g}$ such that $\exp(g) = T$ and $\exp(tg) \in E$
for all $t \in [0,1]$.
Now, let $\bS$ be a \emph{closed} subsemigroup of a Lie group $\bG$ 
with Lie wedge $\rL(\bS)$ and let $\langle \exp \rL(\bS) \rangle_S
:= \{\e^{g_1}\cdots\e^{g_n}\;|\; g_i \in \rL(\bS),\, n \in \N\}$
be the subsemigroup generated by $\exp \rL(\bS) \subset \bG$.
Then
\begin{enumerate}
\item[(i)]
$\bS$ is called \emph{Lie subsemigroup} if it is
characterised by the equality
$\bS = \overline{\langle \exp \rL(\bS)\rangle}_S$;
\item[(ii)]
$\bS$ is called \emph{weakly exponential} if 
$\exp \rL(\bS)$ is dense in $\bS$, i.e., if
$\bS = \overline{\exp \rL(\bS)}$;
\item[(iii)]
$\bS$ is called \emph{exponential} if the set $\bS$ is exponential
in the above sense, i.e., if $\bS = {\exp \rL(\bS)}$;
\item[(iv)]
$\bS$ is called {\em locally exponential} if there exists a 
$\unity$"~neighbourhood basis with respect to $\bS$ consisting of
exponential subsets.
\end{enumerate}
The inclusions 
${\exp \rL(\bS)} \subset
\overline{\exp \rL(\bS)} \subset
\overline{\langle \exp \rL(\bS)\rangle}_S$
are obvious.
A Lie wedge $\frak{w}$ is said to be \emph{global} in $\bG$
if there exists a Lie subsemigroup $\bS \subset \bG$ so that 
$\rL(\bS) = \frak{w}$, i.e. 
$\bS = \overline{\langle \exp(\frak{w})\rangle}_S$.

\begin{remark}
For the sake of completeness note that the term Lie subsemigroup is
closely related (with subtle distinctions) to the notions 
of {\em (completely or strictly) infinitesimally generated}
subsemigroups,
which will not be pursued here any further, cf.~\cite{HHL89}.
\end{remark}

\subsection{The Reductive and the Compact Case}
\label{sec:LieS-Cartan}

Based on the classical Cartan decomposition of
reductive Lie groups \cite{Knapp02}, we reformulate a known result
on the existence of global Lie wedges---a setting which does arise
in open quantum systems, cf. Theorem \ref{thm:opensys-Liewedge}
and Corollary \ref{cor:opensys-Liewedge} below. 
We do so by stating a
convenient version of a more general result, cf.~Theorem V.4.57
and Remark V.4.60 in \cite{HHL89}, streamlined here in view of
practical application.

\begin{theorem}
\label{thm:kp-Liewedge}
Let $\bG$ be a closed connected (matrix) Lie group which is
stable under the conjugate transpose inverse, i.e.~which is
invariant under the involution $\Theta: X \mapsto (X^{-1})^\dagger$.
Let $\g = \mathfrak k \oplus \mathfrak p$ be the decomposition
of its Lie algebra into $+1$ and $-1$ eigenspaces of the involution
${\rm D} \Theta(\unity)=: \theta: X \mapsto -X^\dagger$. Then
\begin{enumerate}
\item[(a)] 
the map $\mathfrak p \times \bK \to \bG$, $(p,K) \mapsto \exp(p)K$
with $\bK := \langle \exp \mathfrak k\rangle$ is a diffeomorphism
onto $\bG$;
\item[(b)] 
the set $\bS := \exp (\mathfrak c) \cdot  \bK$ is a Lie subsemigroup
with $\rL(\bS) =  \mathfrak c \oplus \mathfrak k$, provided
$\mathfrak c \subset \mathfrak p$ is a closed pointed cone,
i.e. $\rE(\mathfrak c) = \{0\}$. 
\end{enumerate}
\end{theorem}


{\bf Proof.}
Combining Proposition~7.14 in 
\cite{Knapp02} with the proof of Theorem~V.4.57 in \cite{HHL89},
the result follows.  \hfill$\blacksquare$

\medskip

Fortunately, the somewhat intricate general scenario just outlined 
simplifies dramatically when considering {\em compact} Lie
subsemigroups.

\begin{proposition}{\bf \cite{HHL89,Lawson99}.}
\label{prop:compactcase}
Let $\bS$ be a compact subsemigroup of a Lie group ${\mathbf G}$.
Then $\bS$ itself is a compact Lie subgroup of ${\mathbf G}$.
\end{proposition}



\subsection{Divisibility and Local Divisibility in Semigroups}
\label{sec:LieS-Div}

Here, we briefly summarise some results on
divisibility in semigroups that will be useful in Section~\ref{sec:Markov-LieS} 
when relating them to recent findings by Wolf~{\em et al.}~on the
divisibility of quantum channels.

For semigroups, there is the following well-established
notion of divisibility~\cite{HHL89,HofRupp91div}: a subset of
$D \subset \bG$ is termed  {\em divisible}, if each element $T \in D$
has roots of any order in $D$, i.e.~to any $r\in\N$ there
is an element $S \in D$ with $S^r = T$.  
Similarly, a semigroup $\bS$ is called {\em locally
divisible}, if there is a $\unity$-neighbourhood basis in $\bS$
consisting of divisible subsets.




For linking global and local notions of divisibility with
exponential semigroups,
Lie semialgebras play a crucial role.
Here we start with some basic results before sketching
what became known as `\/{\em the divisibility problem}\/'.
For details see the literature given in
\emph{Further Notes and References} below.

\begin{proposition}{\bf \cite{HHL89}}
A closed  subsemigroup $\bS$ of a connected Lie group $\bG$ is
divisible if and only if it is exponential, i.e. $\exp \rL(\bS) = \bS$.
\end{proposition}
{\bf Proof.}\hspace{.5em}
If $\bS = \exp \rL(\bS)$, then $\bS$  is trivially divisible.
The converse is already more technical to show
and we refer to Theorem~V.6.5 in \cite{HHL89}.\hfill $\blacksquare$



\begin{theorem}\label{thm:LieSA-locdiv}
For a closed semigroup $\bS$ the following assertions are
equivalent:
\begin{enumerate}
\item[(a)]
the Lie wedge of $\bS$ is a Lie semialgebra;
\item[(b)]
$\bS$ is locally exponential;
\item[(c)]
$\bS$ is locally divisible.
\end{enumerate}
\end{theorem}

{\bf Proof.}
For the equivalence $(a) \Longleftrightarrow (c)$ see
\cite{LNM998} Corollary 3.18 as well as
\cite{HHL89} Propositions IV.1.31-32 and Remark IV.1.14.
While the implication $(b) \Longrightarrow (c)$ is trivial,
$(a) \Longrightarrow (b)$ follows by \cite{LNM998} Proposition 3.17(a).
For a similar result on Lie semigroups see also
\cite{Neeb_LieSG92} Theorem~III.9 and III.21. 
\hfill$\blacksquare$

\medskip


The difficulty to go beyond the straightforward results just mentioned
made the following closely related questions notorious as 
`\/{\em the divisibility problem}\/'
\cite{HHL89, HofRupp91div, HofRupp97div}:
\begin{enumerate}
\item[(i)]
Is the Lie wedge $\mathfrak w = \rL(\bS)$ 
of a closed divisible i.e.~exponential semigroup 
also a Lie semialgebra? 
\item[(ii)] When does (global) divisibility imply local divisibility?
\end{enumerate}

\noindent
These problems were open for several years until settled
in the sterling monography by Hofmann and Ruppert in 1997
\cite{HofRupp97div}, where all Lie groups and subsemigroups
with surjective exponential map are classified. ---
For studying local divisibility in the connected component of the
unity in more detail (and in view of follow-up work), 
some of its main results can be summerised as follows.


\begin{theorem}{\bf \cite{HofRupp97div}}\label{thm:LieW-LieSA}
Let $\bG$ be a connected Lie group containing a weakly exponential
subsemigroup $\bS$ with Lie wedge $\mathfrak w=\rL(\bS)$.
If $\bS$ is closed and has non-empty interior in $\bG$ and its only normal subgroup
is  $\unity\in\bG$, then 
 \begin{enumerate}
 \item[(a)] $\bS$ is divisible (exponential), i.e., $\exp \rL(\bS) = \bS$;
 \item[(b)] its Lie wedge $\mathfrak w =\rL(\bS)$ is a Lie semialgebra; thus
 \item[(c)] $\bS$ is also locally divisible (locally exponential).
 \end{enumerate}
 \end{theorem}

{\bf Proof.} 
For (a) see Theorem 7.3.1 and Scholium 7.3.2 in \cite{HofRupp97div}~(p~132)
lifting Eggert's work \cite{Egg-Diss} on Lie semialgebras to reduced weakly exponential 
subsemigroups thus leading to Theorem~8.2.14 in \cite{HofRupp97div} (p~152);
assertion (b) is Theorem~8.2.1(v) in \cite{HofRupp97div} (p~145); finally
(c) follows from (b) by virtue of Theorem~\ref{thm:LieSA-locdiv} above.
\hspace{.5em} \hfill$\blacksquare$
\medskip


{\em Further Notes and References.} ---
A (somewhat jerry-built) primer on divisible semigroups including an account of earlier
results and problems can be found in \cite{HofRupp91div}, 
while the current status is documented in \cite{HofRupp97div}.
%
A broad overview on historical aspects of a Lie theory of semigroups
is given  in \cite{Lawson92, Hofmann00}. Ultimately, readers interested in
links to Hilbert's Fifth Problem and topological semigroups 
are referred to \cite{Hof94Hilb5}.


\subsection{Reachable Sets}
\label{SSec:Reach}

Let $(\Sigma)$ be a right invariant control system
\begin{equation}
\label{invControlSystem}
\dot{X} = A_u X, \quad A_u \in \g,
\quad u \in \mathcal U \subset \R^m
\end{equation}
on a connected Lie group $\bG$ with Lie algebra $\g$ and let
$\mathfrak{s} \subset \g$ denote its \emph{system Lie algebra},
i.e.~$\mathfrak{s} :=
\langle A_u \;|\; u \in \mathcal U\rangle_{\rm Lie}$
is by definiton the Lie subalgebra generated by $A_u$,
$u \in \mathcal U$.
The \emph{reachable set} $\reach(X_0)$ of ($\Sigma$) is
defined as the set of all $X\in\bG$
that can be reached from $X_0$ by an admissible control
function $u(t)$. More precisely, let $X_u(t)$ denote the
unique solution of Eqn.~(\ref{invControlSystem}) which
corresponds to the control $u(t)$. Then
\begin{equation*}
\reach(X_0) := 
\bigcup_{T \geq 0} \reach(X_0, T)
\end{equation*} 
with
\begin{equation}
\reach(X_0, T):=\{X_u(T)\in\bG~|~ T\geq 0, u(t)\in\mathcal U\}.
\label{reachset}
\end{equation}   
Moreover, $(\Sigma)$ is called \emph{accessible}, if $\reach(X_0)$ has
non-empty interior in $\bG$ for all $X_0 \in \bG$, and
\emph{controllable}, if $\reach(X_0) = \bG$ for all $X_0\in \bG$.
For more details on the control theoretic terminology and setting
we refer to, e.g.,~\cite{JurdKupka81b, Jurdjevic97, Sachkov00}.
Now, in the following series of results the relation between reachable
sets of right invariant control systems and subsemigroups will be
clarified.

\begin{theorem}{\bf\cite{Jurdjevic97, Lawson99}.}
\label{thm:accessibility}
Let $(\Sigma)$  be a right invariant control system on 
$\bG$ given by Eqn.~(\ref{invControlSystem}). Then the
following statements are equivalent:
\begin{enumerate}
\item[(a)] 
The system $(\Sigma)$ is accessible.
\item[(b)] 
The reachable set $\reach(\unity)$ is a subsemigroup of $\bG$
with non-empty interior.
\item[(c)]
The entire Lie algebra $\g$ of $\bG$ is generated by
$A_u, u \in \mathcal U$, i.e. $\mathfrak s = \g$. 
\end{enumerate}
\end{theorem}


\begin{theorem}
\label{thmLiewedge}{\bf\cite{Lawson99}.}
Let $(\Sigma)$ be a right invariant control system on a connected
Lie group $\bG$ given by Eqn.~(\ref{invControlSystem}) and assume
that $(\Sigma)$ is accessible, i.e. $\mathfrak s = \g$.
Then the following statements are satisfied:
\begin{enumerate}
\item[(a)]
The closure of the reachable set $\reach(\unity)$
is a Lie subsemigroup of $\bG$, i.e.
\begin{equation*}
\bS = \overline{\langle \exp \rL(\bS)\rangle}_S
\end{equation*}
where $\bS := \overline{\reach(\unity)}$.
Moreover,
\begin{equation*}
\mathrm{int}\,\bS =
\mathrm{int}\,\big(\reach(\unity)\big),
\end{equation*}
and
\begin{equation}
\label{reachequality}
\bS =
\overline{\reach_{\e}(\unity)},
\end{equation}
where $\reach_{\e}(\unity)$ denotes the reachable
set of the so-called extended system, i.e.~the system
where $A_u$ is allowed to range over the entire Lie wedge
$\rL(\bS)$.
\item[(b)]
The set $\rL(\bS)$ is the largest subset of $\g$ satisfying
(\ref{reachequality}) and, moreover, it is the smallest Lie wedge
which is global in $\bG$ and contains $A_u$, $u \in \mathcal U$.
\end{enumerate}
\end{theorem}


\medskip
\noindent
In control theory, due to the characterisation given in part (b)
of Theorem \ref{thmLiewedge}, the Lie wedge $\rL(\bS)$
is usually known as the \emph{Lie saturate} of $A_u$, $u \in \mathcal U$,
see, e.g., \cite{JurdKupka81a, JurdKupka81b, Kupka90}. 
Conversely, one has the following result.

\begin{theorem}{\bf\cite{Lawson99}.}
\label{thmLiewedgeII}
Let $\bG$ be a connected Lie group and let $\bS$ be a Lie subsemigroup
of $\bG$. Then, there exists a right-invariant control system $(\Sigma)$
on $\bG$ with control set $\{A_u \,|\, u \in \mathcal U\} \subset \g$
such that
\begin{equation*}
\bS := \overline{\reach(\unity)}.
\end{equation*}
In particular, one may choose $\{A_u \;|\; u \in \mathcal U\} = \rL(\bS)$. 
\end{theorem}


\medskip

Finally, we summarise some well-known necessary and sufficient
controllability conditions for  right invariant control
systems. While the first criterion is rather difficult to
check, as the computation of the global Lie wedge corresponding
to a given control set $A_u$ is in general an unsolved problem,
the second one provides a simple algebraic test for compact
Lie groups, cf.~Proposition \ref{prop:compactcase}.

\begin{corollary}
\label{corLiewedge}
Let $(\Sigma)$ be an accessible right invariant control
system on a connected Lie group $\bG$, i.e. $\mathfrak s = \g$. 
Then the following statements are equivalent:
\begin{enumerate}
\item[(a)]
The system $(\Sigma)$ is controllable.
\item[(b)]
The Lie wedge 
of $\overline{\reach(\unity)}$ is all of $\g$.
\end{enumerate}
\end{corollary}

{\bf Proof.} The implication (a) $\Longrightarrow$ (b) is trivial;
the converse (b) $\Longrightarrow$ (a) follows from Theorem
\ref{thm:accessibility}(b) and Theorem \ref{thmLiewedge}(a),
cf. \cite{Lawson99}.
\hfill$\blacksquare$

\begin{corollary}{\bf\cite{Jurdjevic97, JS72}.}
\label{cor:compactcase}
Let $(\Sigma)$ be a right invariant control system on a connected
compact Lie group $\bG$. Then controllability of $(\Sigma)$ is 
equivalent to accessibility, i.e.~to $\mathfrak s = \g$. 
\end {corollary}


\begin{remark}
If the assumption $\mathfrak s = \g$ in Theorem \ref{thmLiewedge}
and Corollary \ref{corLiewedge} is not fulfilled, the above results, however,
still remain valid when restricting to the \emph{unique} Lie group
$\mathbf G_0 := \langle \exp \mathfrak s\rangle$.
\end{remark}


\section{Developments in View of Applications to Quantum Control}


\subsection{Reachable Sets of Closed Quantum Systems}

An application of Corollary \ref{cor:compactcase} to closed
finite-dimensional quantum systems, e.g., $n$ spin-$\frac{1}{2}$
qubit systems with possibly \emph{non-connected}
spin-spin interaction graph yields an explicit characterisation of
their reachable sets. The same result based on a sketchy
controllability argument can be found in \cite{AlbAll02}.

\begin{theorem}
\label{thm:non-connectedgraph}
Assume that the spin-spin interaction graph, which corresponds to
the controlled $n$ spin-$\frac{1}{2}$ system
\begin{equation}
\label{eq:lift}
\dot{U} = -\ri\Big(
H_{d}+\sum_{k=1 \atop \alpha \in \{x,y\}}^{n}u_{k}H_{k,\alpha}\Big) U
\end{equation}
with $H_{d} := \sum_{k < l} J_{kl} \sigma_{k,z}\sigma_{l,z}$
and $H_{k,\alpha} := \sigma_{k,\alpha}$, $\alpha \in \{x,y\}$,
decomposes into $r$ connected components with $n_j$
vertices in the $j$-th component.
Then, the reachable set $\reach(\unity_{2^n})$ of Eqn.~(\ref{eq:lift}) is given
(up to renumbering) by the Kronecker product
$SU(2^{n_1}) \otimes \cdots \otimes SU(2^{n_r})$.   
\end{theorem}

{\bf Proof.} 
Suppose that the spin-$\frac{1}{2}$ particles of the system
are numbered such that the first component of the graph contains the
vertices $1, \dots, n_1$, the second one the vertices
$n_1+1, \dots, n_1+n_2$ and so on. Thus $n = n_1 + \dots + n_r$.
Then, it is straightforward to show that the system Lie algebra
is equal to the Lie algebra of
$\bG_0 := SU(2^{n_1}) \otimes \cdots \otimes SU(2^{n_r})$ cf.~\cite{AlbAll02}.
Therefore, we can consider Eqn.~\eqref{eq:lift} as a control
system on $\bG_0$. Since $\bG_0$ is a closed subgroup of $SU(2^n)$,
it is compact and thus Corollary \ref{cor:compactcase} applied
to $\bG_0$ yields the desired result.
\hfill$\blacksquare$

\medskip
\noindent
Henceforth read $N := 2^n$ for $n$ spin-$\tfrac{1}{2}$ qubits. ---
Note that the same line of argument as above applies to the modified
control term discussed in \cite{AlbAll02}.


\subsection{Open Quantum Systems and Completely
Positive Semigroups}\label{sec:CP-LieS}

In open relaxative quantum systems \cite{Davies76,AlickiLendi87,
Weiss99,BreuPetr02,LNM1880} however, the situation is different because
relaxation translates into `\/contraction\/'. Thus the dynamics
on density operators is no longer described by the action of a
{\em compact} unitary Lie group as before. 

Moreover, we use the following short-hand for the total
Hamiltonian
\begin{equation}
\label{eqn:H_defs}
H_u:= H_d + \sum_j u_j H_j,
\end{equation}
where $u_j$ and $H_j$ denote possibly time dependent control
amplitudes and time-independent control Hamiltonians, respectively. 
Now, we consider a finite dimensional controlled
\emph{Master equation} of motion
\begin{equation}
\label{eqn:cmaster1}
\dot \rho = -\ri \ad{H_u}(\rho) - \Gamma(\rho)
= - {\cal L}_u(\rho),
\quad u \in \mathcal U \subset \R^m
\end{equation}
on the set of density operators
\begin{equation*}
\mathfrak{pos}_1(N) :=  \{\rho \in \gl(N,\C) \,|\, 
\rho = \rho^{\dagger},\, \rho \geq 0,\, \tr \rho =1\}
\end{equation*}
modelling a finite dimensional relaxative quantum system.
Here, $\ad{H_u}$ denotes the adjoint operator,
i.e.~$\ad{H_u}(\rho) := [H_u,\rho]$, 
and $-\Gamma$ represents the infinitesimal generator
of a semigroup $\{\exp(-t \Gamma) \;|\; t\geq 0\}$ of
linear trace- and positivity-preserving
(super-)operators~\footnote{In abuse of language, it is common
to call a positivity-preserving (super-)operator, i.e.~an operator
which leaves the set of positive semidefinite elements
in $\mathfrak{her}(N)$ 
invariant, \emph{positive} for short.}.
Clearly, ${\cal L}_u$ and thus Eqn.~\eqref{eqn:cmaster1}
extend to the vector space of all Hermitian matrices
\begin{equation*}
\mathfrak{her}(N) :=  \{H \in \gl(N,\C) \;|\; 
H = H^{\dagger}\}.
\end{equation*}
Now it makes sense to ask for the self-adjointness of 
$\Gamma$ with respect to the Hilbert-Schmidt inner product
$\langle H_1, H_2 \rangle := \tr (H_1 H_2)$ on $\mathfrak{her}(N)$. Unfortunately,
$\Gamma$ need not be self-adjoint, yet it is self-adjoint,
e.g., if it can be written in double-commutator form,
cf.~Eqn.~\eqref{eqn:doublebracket}.



%
%

Moreover, since the flow of Eqn.~\eqref{eqn:cmaster1} is trace
preserving, the image of $\Gamma$ is contained in the space of
all traceless Hermitian matrices
\begin{equation*}
\mathfrak{her}_0(N) :=  \{H \in \gl(N,\C) \;|\; 
H = H^{\dagger}, \tr H = 0\}.
\end{equation*}
Therefore, the restriction of $\Gamma|_{\mathfrak{her}_0(N)}$
yields an operator from $\mathfrak{her}_0(N)$ to itself 
and thus Eqn.~\eqref{eqn:cmaster1} can also be regarded as
an equation on  $\mathfrak{her}_0(N)$. To distinguish these
two interpretations of Eqn.~\eqref{eqn:cmaster1},
we call the latter \emph{homogeneous Master equation}
\footnote{Note that the term \emph{homogeneous Master equation}
is used here in a general sense and {\em without} any restriction
to high-temperature approximations~\cite{LE95} to Eqn.~\eqref{eqn:cmaster1}.
}. 
%
Note that the homogeneous Master equation completely
characterises the dynamics of the open system, once an equilibrium
state $\rho_*$ of Eqn.~\eqref{eqn:cmaster1} is known.
More precisely, if ${\cal L}_u(\rho_*) = 0$ for all $u \in \R^m$
(e.g.,~choose $\rho_* = \tfrac{1}{N}\unity_N$
for unital equations) the dynamics of
$\rho_0 := \rho - \rho_*$ is described by the
homogeneous Master equation.
Finally, we associate to Eqn.~\eqref{eqn:cmaster1} a
\emph{lifted Master equation}
\begin{equation}
\label{eqn:lift}
\dot{X} = -{\cal L}_u \circ X,
\quad X(0) = \mathrm{id}
\end{equation}
on $GL(\mathfrak{her}(N))$ and $GL(\mathfrak{her}_0(N))$, respectively.
Equation \eqref{eqn:lift} will play a key role in the subsequent
subsemigroup approach.

For a constant control $u(t) \equiv u$, the formal solution of
the lifted Master equation Eqn.~\eqref{eqn:lift} is given by
$T_u(t) := \exp(-t {\cal L}_u)$. Thus
\begin{equation}
\label{eq:masterI}
\{T_u(t)  \;|\; t\geq 0\,\}
\end{equation}
yields a one-parameter semigroup of linear operators acting on
$\mathfrak{her}(N)$. Actually, the operators $T_u(t)$ form a
{\em contraction semigroup of positive and trace preserving
linear operators}
on $\mathfrak{her}(N)$ in the sense that
\begin{equation*}
|| T_u(t)(A)||_{1} \leq
|| A ||_{1}
\end{equation*}
for all $A\in\mathfrak{her}(N)$, cf.~\cite{Koss72,Koss72b}.
Recall that the trace norm $|| A ||_{1}$ of $A\in\mathfrak{her}(N)$
is given by 
\begin{equation*}
||A||_{1}:=\sum_{i}^N \sigma_i = \sum_{i}^N |\lambda_i| \,,
\end{equation*}
where $\sigma_i$ and $\lambda_i$ denote the singular values and 
eigenvalues of $A$, respectively. The semigroup \eqref{eq:masterI} 
is said to be {\em purity-decreasing} if moreover all
$T_u(t)$ constitute a contraction with respect to the norm induced
by the Hilbert-Schmidt inner product, i.e.~if
\begin{equation*}
\big\langle T_u(t)(\rho),T_u(t)(\rho) \big\rangle \leq
\langle \rho,\rho \rangle
\end{equation*}
holds for all $\rho\in\mathfrak{pos}_1(N)$ and all $t \geq 0$.
In general, $T_u(t)$ is \emph{not} purity-decreasing. 
However, if $\Gamma$ is in Kossakowski-Lindblad form,
cf.~Eqn.~\eqref{eqn:LindKoss1}, a necessary and sufficient
condition for being purity-decreasing is unitality of
$\Gamma_L$, i.e.~$\Gamma_L(\unity_N) = 0$, cf.~\cite{Lidar06}.
Thus for a unital Kossakowski-Lindblad term $\Gamma_L$,
the subsemigroup
\begin{equation}
\label{eq:masterII}
{\mathbf P}_\Sigma := 
\langle T_u(t) \;|\; t\geq 0,\, u \in \mathcal U \rangle_S
\end{equation}
generated by the one-parameter semigroups \eqref{eq:masterI}
is contained in a \emph{linear contraction semigroup}
of a Hilbert space.

\begin{remark}
\label{rem:contract}
Let $\mathcal H$ be a complex Hilbert space with scalar product
$\langle \cdot,\cdot \rangle$. Then the
{\em linear contraction semigroup} of $\mathcal H$
is defined by 
\begin{eqnarray*}
\label{eqn:contract}
{\mathbf C(\mathcal H) :=}
\{T \in GL(\mathcal H) \,|\,
\langle Tv,Tv \rangle \leq \langle v,v \rangle\;
\mbox{for all  $v\in \mathcal H$}\}.
\end{eqnarray*}
Note that $\mathbf C_0(\mathcal H)$---the connected component of the
unity in $\mathbf{C}(\mathcal H)$---is in fact a Lie subsemigroup.
This is evident from the polar decomposition $T=PU$, because
$PU \in \mathbf{C}(\mathcal H)$ with
$U$ unitary and $P=P^\dagger$ positive definite holds, if and
only if the eigenvalues of $P$ are at most equal to 1. Thus
\begin{equation*}
\mathbf C_0(\mathcal H) = \exp(- \mathfrak{c})\cdot U(\mathcal H)\,,
\end{equation*}
where $\mathfrak{c}$ denotes the cone of all positive
semidefinite elements in $\gl(\mathcal H)$ and $U(\mathcal H)$
the corresponding unitary group. Similarly, one can define
contraction semigroups for real vector spaces, cf.~\cite{LNM1552}.
\end{remark} 


Next, we briefly fix the fundamental notion of complete
positivity for open quantum systems.
Recall that a linear map $T_u(t)$ is \emph{completely positive},
if $T_u(t)$ and all its extensions of the form
$T_u(t) \otimes \unity_m$ are positivity-preserving, i.e.
\begin{equation*}
\big(T_u(t) \otimes \unity_m\big)
\big(\mathfrak{pos}_1(N \!\cdot\! m)\big)
\subset \mathfrak{pos}_1(N \!\cdot\! m)
\end{equation*}
for all $m \in \N$. Complete positivity of the Markovian 
semigroup $\{T_u(t) \;|\; t \geq 0\}$ is required to guarantee
that $\{T_u(t) \;|\; t \geq 0\}$ can be associated with a
Hamiltonian evolution on a larger Hilbert space,
cf. \cite{Davies76, Rodriguez08, Lidar08}.

According to the celebrated work by Kossakowski \cite{GKS76} and
Lindblad \cite{Lind76}, Eqn.~(\ref{eqn:cmaster1})
generates a one-parameter semigroup $\{T_u(t) \;|\; t \geq 0\}$
of linear trace-preserving and completely positive operators,
if and only if $\Gamma_L$ can be written as
\begin{equation}
\label{eqn:LindKoss1}
\tfrac{1}{2}\sum_{k}
V_k^{\dagger} V_k \rho + \rho V_k^{\dagger} V_k
- 2V_k \rho V_k^{\dagger} =: \Gamma_{L}(\rho) 
\end{equation}
with arbitrary complex matrices $V_k \in \gl(N,\C)$. 
Thus the Master equation (\ref{eqn:cmaster1}) then specialises 
to the \emph{Kossakowski-Lindblad form}
\begin{equation}
\label{eqn:master2}
 {\cal L}_u(\rho)\; := \; \ri\ad{H_u}(\rho)  + \tfrac{1}{2}\sum_{k}
V_k^{\dagger} V_k \rho + \rho V_k^{\dagger} V_k
- 2V_k \rho V_k^{\dagger}.
\end{equation}

\medskip
Suppose we consider the complexification of $\mathfrak{her}(N)$,
i.e.~the complex vector space
$$
\mathfrak{her}(N)^{\C}=\gl(N,\C)= \C^{N \times N}\cong\C^{N^2}.
$$
By extending the linear operators
$\ad{H_u}, \Gamma_{L} \in\gl(\mathfrak{her}(N))$ to
$\widehat H_u,\widehat\Gamma_{L} : \C^{N^2}\rightarrow\C^{N^2}$
one arrives at the superoperator representations
\begin{eqnarray}
\label{eqn:ad}
\widehat H_u &:=&\unity_N \!\otimes\! H_u - H_u^\top\!\otimes\! \unity_N\quad\text{and}\\
\label{eqn:LindKoss2}
\widehat\Gamma_{L} &:=& 
\tfrac{1}{2} \sum_{k=1}^{N^2}\!
\unity_N \!\otimes\! V_k^{\dagger}V_k^{\phantom\dagger}
\!+ V^{\top}_k{V}^*_k \!\otimes\! \unity_N
\!- 2\, {V}^*_k \!\otimes\! V_k^{\phantom *}\,,\qquad
\end{eqnarray}
where $\widehat H_u, \widehat\Gamma_{L}\in\gl(N^2,\C)$ are
$N^2\times N^2$ complex matrices. In particular, if $\Gamma_{L}$ 
is self-adjoint, the corresponding matrix representation
$\widehat\Gamma_{L}\in\gl(N^2,\C)$ is Hermitian.
Moreover, note that the matrix representation $\widehat\Gamma_{L}$
contains some redundancies on $\gl(N^2,\C)$ since the original
$\Gamma_{L}$ operates on the real vector space $\mathfrak{her}(N)$
which has obviously smaller (real) dimension than $\C^{N^2}$.
Viewed in this way, note that $\widehat\Gamma_{L}$
is not the same as the matrix representation of $\Gamma_{L}$
in the \emph{coherence-vector formalism}.
See \cite{AlickiLendi87} for an introduction on coherence vectors in 
open systems and \cite{ByrdKhan03}
for a recent characterisation of positive semidefiniteness in terms of 
Casimir invariants. More geometric features can be found in \cite{SchiZhLea04}.

Now, the previous semigroup theory allows to interpret the
Kossakowski-Lindblad master equation in terms of a Lie wedge
condition.
We define $\bP$ to be the semigroup of all
positive, trace preserving \emph{invertible} linear operators on
$\mathfrak{her}(N)$, i.e. 
\begin{equation*}
\bP := \big\{T \in GL\big(\mathfrak{her}(N)\big)
\;\big|\; T\cdot\mathfrak{pos}_1(N) \subset\mathfrak{pos}_1(N)\big\}.
\end{equation*}
and $\bP^{\rm cp}$ to be the closed subsemigroup of all completely
positive ones, i.e. 
\begin{equation*}
\bP^{\rm cp} := \{T \in \bP ~|~T\text{ completely positive}\}
\subsetneq\bP.
\end{equation*}
Then, $\bP_0$ and $\bP^{\rm cp}_0$ denote the corresponding
connected components of the unity. Moreover, an arbitrary 
linear trace preserving completely positive, not necessarily
invertible operator on $\mathfrak{her}(N)$ is usually called
a \emph{quantum channel}. Thus in terms of quantum channels,
$\bP^{\rm cp}$ is the set of all invertible quantum channels.
Now, a key-result by Kossakowski and Lindblad can
be formulated as follows. 

\begin{theorem} {\bf (Kossakowski, Lindblad \cite{GKS76,Lind76})}
\label{Thm:LKsem}
The Lie wedge $\rL(\bP^{\rm cp}_0)$
is given by the set of all linear operators $- \cal L$ of the form
$ {\cal L} := \ri\adr_H + \Gamma_L$,
where $\Gamma_L$ is defined by Eqn.\eqref{eqn:LindKoss1}.
\end{theorem}

\noindent
While the finite-dimensional version of Theorem
\ref{Thm:LKsem} stated above was originally proven by Gorini,
Kossakowski and Sudarshan \cite{GKS76}, at the same time Lindblad \cite{Lind76}
handled the explicitly infinite-dimesional case of a
norm (uniform) continuous semigroup of completely positive
operators acting on a $W^*$-algebra.
(Note that Kossakowski-Lindblad-type equations with
{\em time dependent} coefficients were analysed, e.g.,~by
\cite{Lendi86} or \cite{CheGarQue97}.)

For proving Theorem \ref{Thm:LKsem}, a former, actually infinite-dimensional 
result by Kossakowski \cite{Koss72} on one-parameter
semigroups of positive (not necessarily completely positive)
operators on trace-class operators $\mathcal B_1(\mathcal H)$ and their
infinitesimal generators was recast into a finite-dimensional 
setting in \cite{GKS76}.
Although Kossakowski and Lindblad exploited different methods
from functional analysis, a crucial point
in both papers \cite{Koss72} and \cite{Lind76} is 
the theory of dissipative semigroups on Banach spaces,
cf.~Lumer and Phillips \cite{LumerPhillips61}.

Yet in the context of finite-dimensional Lie semigroups, 
the same results now show up as
a consequence of a more general invariance theorem for convex cones:
roughly spoken the infinitesimal generator of a one-parameter
semigroup leaving a fixed convex cone invariant is characterised
via its values at the extreme points of the cone,
cf.~Theorem I.5.27 in \cite{HHL89}.
In particular, Kossakowski's work \cite{Koss72} on one-parameter
semigroups of positive operators
then turns out to be a special application of the afore-mentioned
invariance theorem to the convex cone of all positive semidefinite
$N \times N$-matrices
\begin{equation*}
\mathfrak{pos}(N) :=\{H\in\mathfrak{her}(N)~|~H\geq 0\}\quad.
\end{equation*}
Likewise,
Theorem \ref{Thm:LKsem} can be obtained by the invariance
theorem applied to the cone $\mathfrak{pos}(N^2)$, once the
equivalence of complete positivity of $\exp(-t {\cal L})$ and
positivity of $\exp (-t\, {\cal L} \otimes I_N)$ is established,
cf.~\cite{GKS76}. For more details see \cite{Diss-Indra}.


\subsection{{Lie Properties of Semigroups {\em versus}\\ Markov Properties of Quantum Channels}}
\label{sec:Markov-LieS}

Recall the notation $\bP^{\rm cp}$ for the closed semigroup
of all completely positive invertible maps, whose connected component
of the unity is termed $\bP^{\rm cp}_0$.
Having derived the Lie wedge of $\bP^{\rm cp}_0$,
the issue of its {\em globality}
naturally emerges. Since $\bP^{\rm cp}_0$ is closed in $GL(\mathfrak{her}(N))$,
an affirmative answer to this problem is obtained by Proposition V.1.14
in \cite{HHL89}.

\begin{theorem}
\label{Thm:maxraysg}
The semigroup
\begin{equation}\label{eqn:LieSG-Pcp0}
\bT := \overline{\big\langle\exp\big(\rL(\bP^{\rm cp}_0)\big)\big\rangle}_{S}
\subseteq\bP^{\rm cp}_0
\end{equation}
generated by $\rL(\bP^{\rm cp}_0)$ is
a Lie subsemigroup with the Lie wedge
$\rL(\bT)=\rL(\bP^{\rm  cp}_0)$. In particular, $\rL(\bP^{\rm cp}_0)$
is a global Lie wedge.
\end{theorem}

Ultimately, the question arises whether $\bP^{\rm cp}_0$ is itself a {\em Lie}
subsemigroup in the sense of Section \ref{sec:basics}.
However, the identity $\bT=\bP^{\rm cp}_0$ one might surmise is disproven
by the fact that there are indeed invertible quantum channels $T$
with $\det T > 0$ that do not belong to the subgroup $\bT$,
cf.~\cite{Wolf08a,Wolf08b}.

For relating these references to our context, we
have to establish some of the terminology of 
Holevo~\cite{Holevo01} and Wolf~{\em et al.}~\cite{Wolf08a,Wolf08b}:
Similar to our definition in Section \ref{sec:LieS-Div}, a quantum channel
$T$ is called {\em (inifinitely) divisible}
if for all $r \in \N$ there exists a channel $S$ such
that \mbox{$T = S^r$}. 
  [NB: In stochastics and quantum physics \cite{Holevo86,Denisov88,Holevo01,Wolf08a,Wolf08b} 
  it is long established to
  use the term `\/infinitely divisible\/', whereas in mathematical semigroup theory
  it is equally long established to simply say `\/divisible\/' instead
  (see also Section~\ref{sec:LieS-Div}). 
  This is why here we use the brackets.] 
In contrast, a channel is said to be {\em infinitesimal divisible}
if for all $\varepsilon > 0$ there is a sequence of channels
$S_1, S_2, \dots, S_r$ such that $\|S_j - \mathrm{id}\| \leq \varepsilon$
and $\prod_{j=1}^r S_j=T$. Moreoever, a quantum channel
is termed {\em time (in)dependent Markovian} if it is the solution of a
Master equation $\dot{X} = -{\cal L} \circ X$,
with initial condition
$X(0) = \mathrm{id}$ and {\em time (in)dependent
Liouvillian} $-{\cal L}$ of Kossakowski-Lindblad form.
Now, for our purpose the results in \cite{Wolf08a,Wolf08b} can be
resumed as follows.

\begin{proposition}{\bf \cite{Denisov88, Wolf08a}}\label{prop:wolf08a}
\begin{enumerate}
\item[(a)]
The set of all time independent Markovian channels coincides with the 
set of all (infinitely) divisible and invertible channels. 
\item[(b)]
The closure of the set of all time dependent Markovian channels
coincides with the closure of the set of all infinitesimal divisible
channels.
\end{enumerate}
\end{proposition}

\noindent
The proof of Proposition \ref{prop:wolf08a} (a) is given in 
\cite{Denisov88}, part (b) is precisely Theorem~16 of 
\cite{Wolf08a}.
Thus in relation to the work of Wolf {\it et al.} Theorem~\ref{Thm:maxraysg}
reads:

\begin{corollary}
The closure of the set of all time dependent Markovian channels
forms the Lie subsemigroup $\bT$ defined in \eqref{eqn:LieSG-Pcp0}. 
Its tangent space at the unity is given by the {\em Lie wedge}
$L(\bP^{\rm cp}_{\rm 0})$ of all Kossakowski-Lindblad generators.
\end{corollary}

\noindent
However, one also arrives at the no--go result:

\begin{theorem}{\bf \cite{Wolf08a}}\label{thm:wolf08a}
The semigroup $\bP_0^{\rm cp}$ is neither (infinitely) divisible nor infinitesimal
divisible. In particular, there are invertible quantum channels which
are not infinitesimal divisible.
\end{theorem}

\noindent
For $N =2$, the above assertion is rigorously proven by Theorem~24
in \cite{Wolf08a}. For $N>2$, the statement currently presupposes
one may extrapolate
from the numerical results (also for $N=2$) in \cite{Wolf08b}.

Now, from Theorem \ref{thm:wolf08a} we conclude:
\begin{corollary}\label{cor:wolf08a}
$\bP^{\rm cp}_0$ itself is not a Lie subsemigroup.
Yet in particular 
the semigroup $\bP^{\rm cp}$ of all invertible quantum channels
is made of three subsets, all of which also occur in the connected 
component $\bP^{\rm cp}_0$: 
\begin{enumerate}
\item[(a)]
the set of time independent Markovian channels which is given by
definition as the union of all one-parameter Lie semigroups
$\{\exp(-\mathcal L t) \;|\; t \geq 0\}$ with $-\mathcal L$ in
Kossakowski-Lindblad form;
\item[(b)]
the closure of the set of time dependent Markovian channels
which coincides with the Lie semigroup $\bT$ defined by
\eqref{eqn:LieSG-Pcp0}\,;
\item[(c)]
besides, there is a set of non-Markovian channels 
(i.e.~neither time independent nor time dependent Markovian)
whose intersection with $\bP^{\rm cp}_0$
has non-empty interior. 
\end{enumerate}
\end{corollary}

Clearly, Markovian channels of type (a) are a special
case of type (b) and (a) is even a proper subset of (b),
since $\bT$ is not exponential \footnote{There are quantum
channels in $\bT$ having pairwise distinct negative eigenvalues.
Such channels are clearly not time independent Markovian, because they
do not have \emph{any} real logarithm in $\gl(\mathfrak{her}(N))$
\cite{Wolf08pc}. }. 
There are also quantum channels with $\det T\leq 0$ \cite{Wolf08a},
but they can only occur outside the connected component
$\bP^{\rm cp}_0$, and thus they are obviously non-Markovian.
The geometry of non-Markovian channels seems to be well-understood
in the single-qubit case ($N=2$), yet remains to be analysed in
full detail for larger $N$.

\begin{corollary}\label{cor:loc-div}
\begin{enumerate}
\item[(a)]
The semigroup $\bP^{\rm cp}_0$ is neither locally divisible
nor locally exponential. 
\item[(b)]
The Lie wedge $L(\bP^{\rm cp}_{\rm 0})$
of all Kossakowski-Lindblad generators does not form a Lie semialgebra.
\end{enumerate}
\end{corollary}

{\bf Proof.} 
Again, for $N=2$, part (a) follows from Theorem~24 in 
\cite{Wolf08a}. For $N>2$, the assertion extrapolates from
the numerical results 
in \cite{Wolf08b}.
Part (b) is an immediate consequence of part (a) and
Theorem~\ref{thm:LieSA-locdiv}.
\hfill$\blacksquare$

\medskip

Now, the distinction between Lie wedge and Lie-semialgebra
structure can be exploited to separate between time dependent
Markovian quantum channels and time {\em in}\/dependent ones.
In general, this separation
is rather delicate. Clearly, as soon as a time
dependent channel $T$ has a representation of the form
$T = \prod_{j=1}^{r}S_j$ such that the 
$S_1, S_2, \dots, S_r$ generate an {\em exponential Lie semigroup}, 
then $T$ is actually time independent.
Though almost a tautology, this statement is quite difficult to check
and therefore an (infinitesimal) condition that is easier to verify is most
desirable.
The following corollary is meant as a first result in this direction---with the
shortcoming that it applies to channels close to unity.
\begin{corollary}
Let $T$ be a time dependent Markovian channel
that allows for a representation $T = \prod_{j=1}^{r}S_j$
with $S_1 = e^{-\mathcal L_1}, S_2 = e^{-\mathcal L_2},
\dots, S_r = e^{-\mathcal L_r}$
and where $\mathfrak w_r$ denotes the smallest global Lie wedge
generated by ${\mathcal L_1}, {\mathcal L_2},\dots, {\mathcal L_r}$.
Then
\begin{enumerate}
\item[(a)] $T$ boils down to a time {\em in}\/dependent Markovian channel,
           if it is suffiently close to the unity and 
           if there is a representation so that the associated Lie wedge $\mathfrak w_r$ 
           also carries Lie-semialgebra structure;
\item[(b)] conversely, if $T$ is a time {\em in}\/dependent Markovian channel, a
           representation with $\mathfrak w_r$ being a Lie semialgebra trivially exists.
\end{enumerate}
\end{corollary}

{\bf Proof.} 
The result follows by the same line of arguments
as Corollary~\ref{cor:ALT-LieSA} below. 
\hfill$\blacksquare$

\medskip

\medskip

Thus in summary three elucidating results have emerged:
(i) the set of all time dependent Markovian quantum channels forms
a {\em Lie subsemigroup} $\bT$ and (ii)~its {\em Lie wedge} coincides with 
with the set of all Kossakowski-Lindblad operators: it is the
Lie wedge to the subsemigroup $\bP^{\rm cp}_0$ of all
invertible quantum maps.
Moreover, (iii) the border from time dependent to time independent
Markovian quantum channels is characterised by
the existence of an associated Lie wedge that
specialises to {\em Lie-semialgebra} structure.

\subsection{Effective Liouvillians}

In physical applications a
frequent task amounts to  describing the evolution of a 
controlled Master equation
\begin{equation}
\label{eq:cmaster}
\dot X = - {\cal L}_u \circ X,
\quad u \in \mathcal U \subset \R^m,
\end{equation}
cf.~Eqns.~(\ref{invControlSystem},~\ref{eqn:cmaster1}) with $\mathcal L_{u}$
in Kossakowski-Lindblad from Eqn.~\eqref{eqn:master2},
by an appropriate one-parameter semigroup.
More precisely, given an admissible time dependent control
$u(t) \in \mathcal U$ and a final time $t_{\rm eff} > 0$ one
is interested in an {\em effective time-independent Liouvillian}
$\mathcal L_{\rm eff}$ such that the two time evolutions coincide
at $t_{\rm eff} > 0$, i.e.
$T_{u(t)}(t_{\rm eff}) = e^{-t_{\rm eff} \mathcal L_{\rm eff}}$.
This is a natural extension from average Hamiltonian theory of
closed systems to average Liouvillians of open ones
\cite{Magnus54, Maricq90, Ghose00, Casas07}.

Now, Lie-semigroup theory provides a useful framework to settle the
question under which conditions not only the final point
$e^{-t_{\rm eff}\mathcal L_{\rm eff}}$,
but also the {\em entire trajectory}
$\{e^{-t \mathcal L_{\rm eff}}|0\leq t \leq t_{\rm eff}\}$
up to the final point complies with the Master equation \eqref{eq:cmaster}
defining the physics of the system.

\begin{corollary}\label{cor:ALT-LieSA}
Given a Master equation \eqref{eq:cmaster}
and the smallest global Lie wedge $\mathfrak{w}$ generated by the set
of controls $\{-\mathcal L_u\,|\,u\in \mathcal U \subset \R^m\}$, 
cf.~Theorem~\ref{thm:accessibility}.
Then the following assertions are equivalent:
\begin{enumerate}
\item[(a)]
The Lie wedge $\mathfrak{w}$ also is a Lie semialgebra.
\item[(b)] 
Any solution of  \eqref{eq:cmaster}
coincides at least locally, i.e.~for sufficiently small $t > 0$ with
some one-parameter semigroup generated by
an effective Liouvillian $\mathcal L_{\rm eff} \in \mathfrak{w}$.
\end{enumerate}
\end{corollary}

{\bf Proof.} Follows from the fact that the Lie semigroup
$\overline{\langle\exp \mathfrak{w}\rangle}_S$ is
{\em locally exponential} if and only if its Lie wedge is a
Lie semialgebra, cf.~Theorem~\ref{thm:LieSA-locdiv}.
\hfill$\blacksquare$

\medskip

Only if the effective Liouvillian is guaranteed to remain
within  the Lie wedge $\mathfrak{w}$ associated to the controlled
Master equation \eqref{eq:cmaster} then it generates a one-parameter
semigroup  $\{e^{-t\mathcal L_{\rm eff}}\,|\, t \geq 0\}$ that can be
considered \/`physical\/' at {\em all times} $t > 0$.
Otherwise, the physical validity of the time evolution described by
the semigroup $\{e^{-t \mathcal L_{\rm eff}}|t \geq 0\}$ is in general
limited to a set of discrete times (including $t=0$ and $t=t_{\rm eff}$).

\subsection{Controllability Aspects of Open Quantum Systems}

\subsubsection*{Structural Preliminaries}

Studying reachable sets of open quantum systems subject to a controlled
Hamiltionian, cf.~Eqn.~\eqref{Eq:CLKE} below, is intricate, as will be evident
already in the following simple scenario: consider a Master equation
in the superoperator form
$$
\vec \dot \rho = -(\ri \sum_j\widehat H_j +\widehat\Gamma_L)\vec \rho\,,
$$
where the $\ri \widehat H_j$ are skew-Hermitian, while $\widehat\Gamma_L$ 
shall be Hermitian.
Thus they respect the standard Cartan decomposition of
$\mathfrak{gl}(N^2,\C{}) := \mathfrak k \oplus \mathfrak p$
into skew-Hermitian matrices ($\mathfrak k$) and Hermitian matrices
($\mathfrak p$). Then the usual commutator relations
$[\mathfrak k, \mathfrak k]\subseteq \mathfrak k, \;
[\mathfrak p, \mathfrak p]\subseteq \mathfrak k, \;
[\mathfrak k, \mathfrak p]\subseteq \mathfrak p$
suggest that double commutators of the form
$$
\big[[\widehat H_j,\widehat \Gamma_L],
[\widehat H_k,\widehat \Gamma_L]\big]
$$
generate \emph{new} $\mathfrak k$-directions in the system
Lie algebra as will be described below in more detail.

For the moment note on a general scale that 
such controlled open systems thus fail to comply with
the standard notions of controllability: not only does this hold for
operator controllability of the lifted system
but also for usual controllability on the set 
of all density operators,
cf. \cite{DiHeGAMM08, Alt03}. Hence it is natural to ask for
weaker controllability concepts in open systems.

For simplicity, we confine the subsequent considerations to 
\emph{unital} systems of Kossakowski-Lindblad form, 
i.e. $\Gamma_{L}(\unity_N) = 0$, as their dynamics is
completely described by the homogeneous Master equation
\begin{equation}
\label{Eq:CLKE}
\begin{split}
\dot \rho &= -\ri \adr_{H_u}(\rho) - \Gamma_{L}(\rho)=
-\mathcal{L}_u(\rho)
\end{split}
\end{equation}
on $\mathfrak{her}_0(N)$ and its lift
\begin{equation}
\label{eq:masterIII}
\begin{split}
\dot X &= -\mathcal{L}_u \circ X
\end{split}
\end{equation}
to $GL(\mathfrak{her}_0(N))$. 
Here the controlled Hamiltionian takes the form
of Eqn.~\eqref{eqn:H_defs} 
with $ H_d$ and  $H_j$ in $\su(N)$ and \emph{no bounds} on
the controls $u_j \in \R$. Thus the semigroup $\bP_\Sigma$ given
by Eqn.~\eqref{eq:masterIII} will be regarded as a subsemigroup
of $GL(\mathfrak{her}_0(N))$ in the sequel. Alternatively,
by the previously introduced superoperator representation,
we can think of $\bP_\Sigma$ as embedded in $GL(N^2,\C)$.

If, in the absence of relaxation, the Hamiltonian system
is fully controllable, we have 
\begin{equation}
\label{eqn:fullcontrol}
\langle \ri H_d,\ri H_j\;|\; j=1,\dots,m\rangle_{\sf Lie} =
\mathfrak{su}(N) \;,
\end{equation}
or, equivalently,
\begin{equation*}
\langle \ri\widehat H_d,\ri\widehat H_j\;|\; j=1,\dots,m\rangle_{\sf Lie}
= \mathfrak{psu}(N) \subset \mathfrak{su}(N^2)\;,
\end{equation*}
where we envisage $\mathfrak{psu}(N)$ to be represented as
Lie subalgebra of $\mathfrak{su}(N^2)$ given by all matices 
of the form $\ri(\unity\otimes H - H^{\top}\otimes\unity)$
with $\ri H \in \mathfrak{su}(N)$. Master equations which
satisfy Eqn.~\eqref{eqn:fullcontrol}
are expected to be generically accessible, i.e.~their system
Lie algebras generically meet the condition 
\begin{equation*}
\langle  \ri\ad{H_d} + \Gamma_L, \ri\ad{H_j}
\;|\; j=1,2,\dots,m \rangle_{\sf Lie} =
\gl(\mathfrak{her}_0(N))\;,
\end{equation*}
cf.~\cite{DiHeGAMM08,Alt04,Diss-Indra}.
Here, the system Lie algebra of the control system (cf. Section
\ref{SSec:Reach}) is not to be misunderstood as its Lie wedge, which
in general is but a proper subset of the system Lie algebra.

The \emph{group} generated by Eqn.~\eqref{eq:masterIII} therefore
generically coincides with $GL(\mathfrak{her}_0(N))$. 
Thus already the coherent part of the open system's dynamics,
i.e.~the `\/orthogonal part\/' of the polar decomposition 
of elements in $\bP_\Sigma$, has to be embedded into a
larger orthogonal (unitary) group than of the same system
being closed, \mbox{i.e.~when $\Gamma_L = 0$.}
This can easily be seen if the Master equation \eqref{Eq:CLKE}
specialises so that the respective matrix representations $\ri \widehat H_j$
for $\ri\adr_{H_j}$ are skew-Hermitian, while $\widehat \Gamma_L$ 
is Hermitian. For instance, this is the case in
the simple double-commutator form
\begin{equation}
\label{eqn:doublebracket}
\begin{split}
\dot \rho &= -\big(\ri \adr_{H_u} 
+ \tfrac{1}{2}\sum_k \adr^2_{V_k}\big) (\rho)\quad.
\end{split}
\end{equation}
It exemplifies the details why
iterated commutators like
$\big[[\widehat H_j,\widehat \Gamma_L],
[\widehat H_k,\widehat \Gamma_L]\big]$
typically generate new
skew-Hermitian directions in the system Lie algebra of
Eqn.~\eqref{eq:masterIII}.
This holds \/{\em a forteriori}\/ if---as henceforth---we allow for
general Kossakowski-Lindblad generators no longer confined to be 
in double-commutator form \eqref{eqn:doublebracket}. We can therefore
summarise the above considerations as follows.

\medskip
\noindent
{\bf Resume.}
{\em In open quantum systems that are fully controllable for
$\Gamma_L= 0$, one finds:
\begin{enumerate}
\item
Only if $\Gamma_L|_{\mathfrak{her}_0(N))}$ acts as scalar $\gamma \unity$
and thus $[\ri H_j, \Gamma_L] = 0$ for all $j$, 
the open dynamics is confined to the contraction semigroup 
$(0,1] \cdot \Adr_{SU(N)}$ of the unitary adjoint group
$\Adr_{SU(N)}$. Moreover, the contractive relaxative part
and the coherent Hamiltonian part are independent in 
the sense that their interference does not generate 
new directions in the
Lie algebra.
\item
Yet in the generic case,
the open systems' dynamics explore a semigroup larger
than the contraction semigroup of the unitary part $\Adr_{SU(N)}$ of the
closed analogue.
\end{enumerate}
}

\medskip
\noindent
Thus for an explorative overview, the task is three-fold: 
\begin{enumerate}
\item[(i)] find the system Lie algebra  
\begin{equation}
\mathfrak s_{\rm open}
:= \langle \ri\ad{H_d} + \Gamma_L, 
\ri\ad{H_j}\rangle_{\sf Lie}\;;
\end{equation}
\item[(ii)]
if $\mathfrak s_{\rm open} = \gl(\mathfrak{her}_0(N))$ already 
(as will turn out to be the case in most of the physical
applications with generic relaxative parts $\Gamma_L$),
then the dynamics of the entire open system takes the form of a
contraction semigroup contained in $GL(\mathfrak{her}_0(N))$;
the relaxative part interferes with the coherent Hamiltonian part
generating new directions in the Lie algebra, where the geometry
of the interplay determines the set of explored states;

\item[(iii)]
in the (physically rare) event of
$\mathfrak s_{\rm open} \subsetneqq \gl(\mathfrak{her}_0(N))$ 
the system dynamics takes the form of a contraction semigroup
contained in a proper subgroup of $GL(\mathfrak{her}_0(N))$.
\end{enumerate}


\subsubsection*{Weak Hamiltionian Controllability}

As mentioned before, controllability notions for open systems
weaker than the standard one are desirable, since Eqn.~\eqref{Eq:CLKE}
is in general non-controllable in the usual sense.
Here, we define a unital open quantum system
to be {\em Hamiltonian controllable}
({\sc h}-controllable) if the subgroup
$\{\Adr_U \,|\, U \in SU(N)\}$ is contained in the
closure of the subsemigroup $\bP_\Sigma$, i.e.  
\begin{equation*}
\{\Adr_U \,|\, U \in SU(N)\} \subset \overline{\bP}_\Sigma.
\end{equation*}
In constrast, we will call a system to be
{\em weakly Hamiltonian controllable} ({\sc wh}-controllable)
if the subgroup $\{\Adr_U \,|\, U \in SU(N)\}$ is contained in the
closure of the subsemigroup
$\R^+ \cdot \mathbf P_{\Sigma} \subset GL(\mathfrak{her}_0(N))$, i.e.  
\begin{equation*}
\{\Adr_U \,|\, U \in SU(N)\} \subset
[1,\infty) \cdot \overline{\bP}_\Sigma.
\end{equation*}
So far, {\sc wh}-controllability has not been studied in the
literature, although it provides a partial answer to the
problem of finding the best approximation to a target density
operator $\rho_F$ by elements of the reachable set $\reach(\rho)$,
where $\rho_F$ itself is contained in the unitary orbit $\mathcal O(\rho)$. 
For establishing a first basic result on {\sc wh}-controllable
systems,
the subalgebras generated by the controls terms
\begin{equation*}
\mathfrak k_{c}
:= \expt{\ri H_{1}, \dots, \ri H_{m}}_{\rm Lie}
\end{equation*}
and by the  \emph{Hamiltionian drift} plus controls terms
\begin{equation*}
\mathfrak k_{d}
:= \expt{\ri H_{d}, \ri H_{1}, \dots, \ri H_{m}}_{\rm Lie}
\end{equation*}
will play an essential role. 

\begin{figure*}[Ht!]
\mbox{\hspace{2mm} \sf (a)\hspace{92mm}(b)\hspace{70mm}}\\
\includegraphics[scale=0.5]{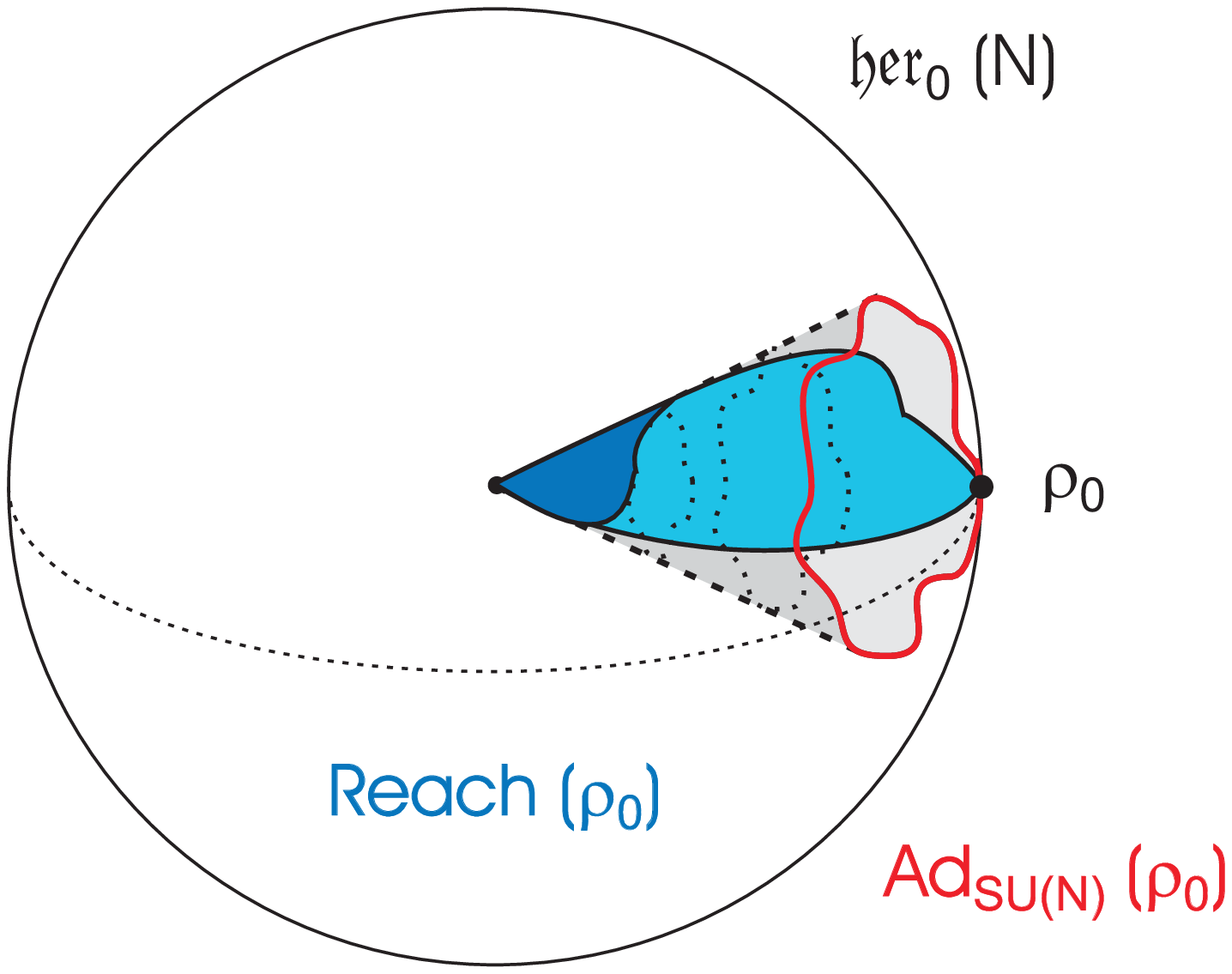}\hspace{22mm}
\raisebox{0mm}{\includegraphics[scale=0.5]{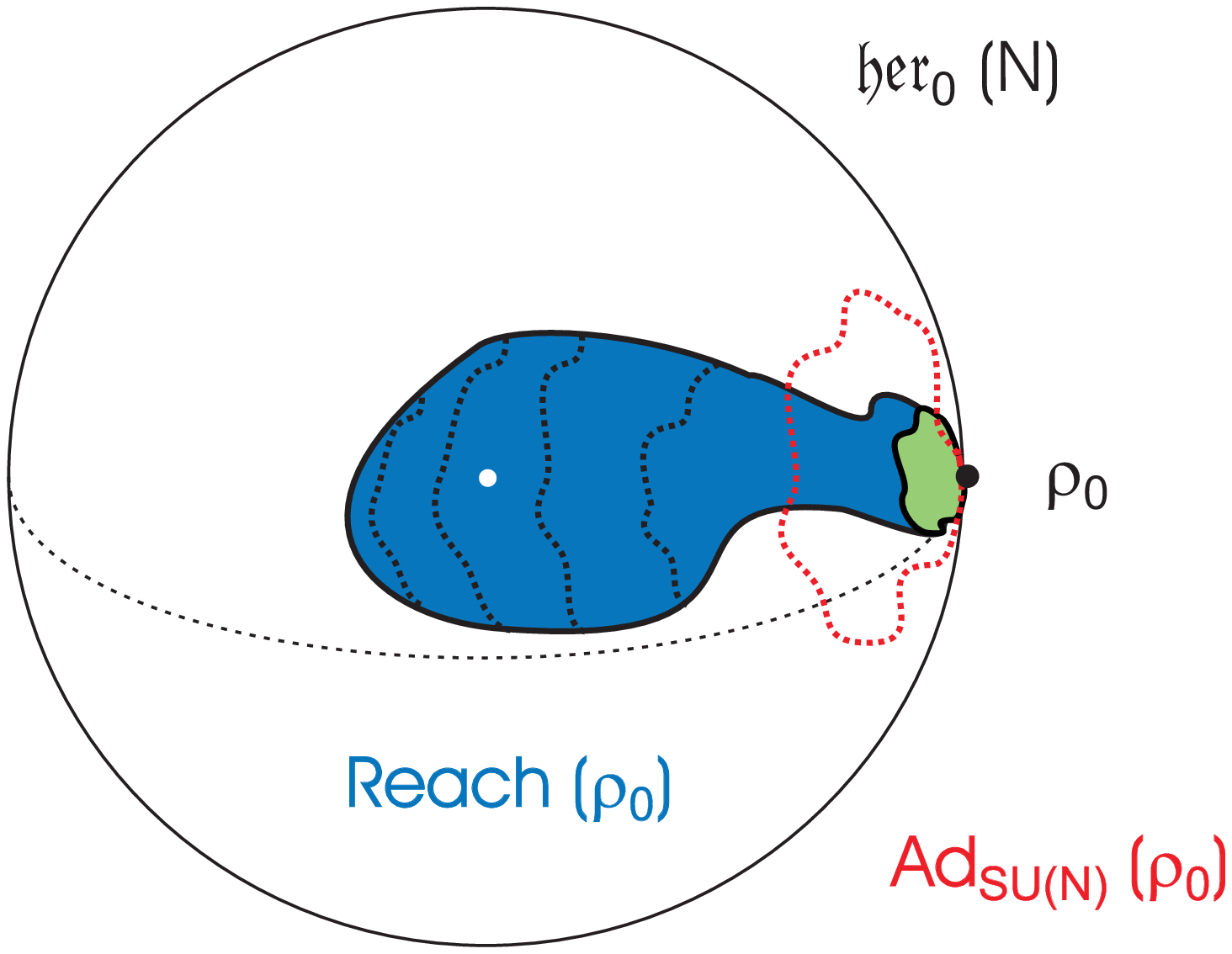}}\\[4mm]
\caption{ \label{fig:reach_WH_g}(Colour online)
Quantum state-space manifolds for open relaxative systems shown as
subsets of $\mathfrak{her}_0(N)$ with scales corresponding to
the metric induced by the Hilbert-Schmidt scalar product.
The centre of the high-dimensional sphere is the zero-matrix,
and the geometry refers to larger systems, e.g., multi-qubit systems with $N\geq 4$.
If in the absence of relaxation, the system
is fully controllable, the reachable set for a fixed initial
state represented as density operator $\rho_0$
takes the form of the entire unitary orbit $\Adr_{SU(N)}(\rho_0)$.
It serves as a reference and is shown as closed curve in red. 
In the text we focus on two different scenarios of open systems:
(a) Dynamics of {\em weakly Hamiltonian controllable} systems with
the Kossakowski-Lindblad term acting approximately as scalar
$\Gamma_L\simeq \gamma\unity$ are confined to the subset (marked in blue)
of states evolving from $\rho_0$ under the action of the contraction 
semigroup $(0,1] \cdot \Adr_{SU(N)}$. The latter is
depicted as grey {\em surface} of a `\/funnel\/' intersecting
the surface of the high-dimensional sphere in the unitary orbit.
Towards the origin, i.e., at long times, the reachable set of 
{\sc wh}-controllable systems typically wraps the entire 
surface (dark blue portion).
(b) In the {\em generic case} when $[\Gamma_L,H_\nu]\neq 0$ ($\nu = d;1,2,\dots, m$),
the dynamics with initial state $\rho_0$ evolves within the {\em volume}
shown in blue. New directions due to the interplay of coherent
Hamiltonian evolution and relaxation make the dynamics explore
a much larger state space than resulting from the simple contraction
semigroup $(0,1] \cdot \Adr_{SU(N)}$, i.e.~the surface in part (a) or 
even the volume contained in its interior.
The intersection (green portion) of the volume $\reach(\rho_0)$ with
the surface of the sphere consists of the set of all states
reachable from $\rho_0$ in zero time or without relaxative loss.
This may often collapse to the single point $\rho_0$ or its {\em local}
unitary orbit \cite{WONRA_tosh,WONRA_dirr}.
}
\end{figure*}

\begin{proposition}
\label{prop:weak-semi-control}
A unital open quantum system \eqref{Eq:CLKE} with
the Hamiltonian given by Eqn.~\eqref{eqn:H_defs} is 
\begin{enumerate}
\item[(a)]
{\sc h}-controllable, if $\mathfrak k_c = \su(N)$ and 
no bounds on the control amplitutes $u_j$, $j = 1, \dots, m$
are imposed;
\item[(b)]
{\sc wh}-controllable, if 
$\mathfrak k_{d} = \su(N)$ and
$\Gamma_L\big|_{\mathfrak{her}_0(N)} \!=\! \gamma \unity$
with $\gamma \geq 0$.
\end{enumerate}
Moreover, for $U \in SU(N)$, the smallest $\lambda \in \R^+$
such that $\Ad{U} \in \lambda \overline{\mathbf P}_\Sigma$
is given by $\e^{\gamma T^*(U)}$,
where $T^*(U)$ denotes the optimal time to steer the lifted system
given by Eqn.~\eqref{eq:masterIII} without relaxation, i.e.~for
$\Gamma_L = 0$, from the identity $\unity$ to $\Ad{U}$.
In particular, for  $\mathfrak k_c = \su(N)$ one has
$\lambda=1$ for all $U \in SU(N)$.
\end{proposition}

{\bf Proof.} (a) First, suppose $\mathfrak k_c = \su(N)$.
Then, for $\Gamma_L = 0$ the fact that we do not assume any
bounds on the controls $u_j \in \R$ implies  that one can steer
from the identity $\unity$ to any $\Ad{U}$ arbitrarily fast. 
Thus for $\Gamma_L \neq 0$ a standard continuity argument
from the theory of ordinary differential equations shows that 
one can approximate $\Ad{U}$ up to any accuracy by elements of
$\mathbf P_\Sigma$. Thus {\sc h}-controllability holds.

\medskip
\noindent
(b) Suppose $\mathfrak k_{d} = \su(N)$ and
$\Gamma_L\big|_{\mathfrak{her}_0(N)} = \gamma \unity$.
By Corollary \ref{cor:compactcase}, we obtain controllability of 
$\{\Adr_U \,|\, U \in SU(N)\}$ for $\Gamma_L = 0$.
Therefore, we can choose a control $u(t)$ which steers the
identity $\unity$ to $\Ad{U}$ in optimal time $T^*(U)$.
Applying the same control to the system under relaxation
yields a trajectory which finally arrives at
$\e^{-\gamma T^*(U)} \Ad{U}$. Thus {\sc wh}-controllability holds
for $\lambda = \e^{\gamma T^*(U)}$. Moreover, by the time optimality
of $T^*(U)$ it is guaranteed that $\lambda = \e^{\gamma T^*(U)}$
is the smallest $\lambda \in \R^+$ such that
$\Ad{U} \in \lambda \overline{\mathbf P}_\Sigma$ holds.
\hfill$\blacksquare$

\medskip

In general, an open quantum system that is fully controllable
in the absence of relaxation will not be necessarily 
{\sc wh}-controllable when including relaxation, even though
it may be accessible. A counterexample showing this fact for
the simplest two-level system and simulations will be provided
in \cite{Diss-Indra}.
Establishing necessary and sufficient conditions for {\sc wh}-controllability
of open quantum systems is therefore an open research problem.
For unital systems which are controllable in 
the absence of relaxation, we
do expect that the `\/ratio\/' of the Hamiltonian and the relaxative
drift term completely determines {\sc wh}-controllability. ---
Finally we will see that additional
assumptions ensuring the preconditions of Theorem
\ref{thm:kp-Liewedge}
allow for inclusion of the global Lie wedge of Eqn.~\eqref{Eq:CLKE}.

\begin{theorem}
\label{thm:opensys-Liewedge}
Assume that the unital Master equation \eqref{Eq:CLKE}
with the Hamiltonian given by Eqn.~\eqref{eqn:H_defs} fulfills
the following condition:
there exists a pointed cone $\mathfrak{c}$ in the set of all 
positive semidefinite linear operators on $\mathfrak{her}_0(N)$
such that 
\begin{enumerate}
\item
$\Gamma_L\big|_{\mathfrak{her}_0(N)} \in \mathfrak{c}$\,;
\item
$[\mathfrak{c},\mathfrak{c}] \subset \ad{\su(N)}$
and 
$[\mathfrak{c},\ad{\su(N)}] \subset \mathfrak{c} - \mathfrak{c}$\,;
\item
$\Ad{U} \mathfrak{c} \Ad{U^{-1}} \subset \mathfrak{c}$
for all $U \in \SU(N)$\,.
\end{enumerate}
Then, the Lie subsemigroup $\overline{\mathbf P}_{\Sigma}$ of
Eqn.~\eqref{Eq:CLKE}
is contained in the Lie subsemigroup
\begin{equation*}
\exp(- \mathfrak c) \cdot \Ad{SU(N)}
\end{equation*} 
with Lie wedge $(-\mathfrak c) \oplus \ad{\su(N)}$.
\end{theorem}

{\bf Proof.} By Theorem \ref{thmLiewedge}(b), it is sufficient
to verify that $\exp(- \mathfrak c) \cdot \Ad{SU(N)}$ is a Lie
subsemigroup with Lie wedge $(-\mathfrak c) \oplus \ad{\su(N)}$.
This will be achieved by applying Theorem \ref{thm:kp-Liewedge}.
To this end, we define $\g := \mathfrak k \oplus \mathfrak p$ with
$\mathfrak k := \ad{\su(N)}$ and 
$\mathfrak p := (\mathfrak c - \mathfrak c) +
(\mathfrak c - \mathfrak c)^{\top}$. Note that the set
$\mathfrak c - \mathfrak c$ consisting of all differences within
the cone $\mathfrak c$ coincides with the vector space spanned by
$\mathfrak c$. Thus $\mathfrak p$ is a subspace of
$\mathfrak{gl}(\mathfrak{her}_0(N))$ which is invariant
under the involution $\Lambda \mapsto - \Lambda^\top$, where
$\Lambda^\top$ denotes the adjoint operator of $\Lambda$ with respect
to the Hilbert-Schmidt inner product on $\mathfrak{her}_0(N)$.
Then, $\g$ constitutes a Lie subalgebra of
$\mathfrak{gl}(\mathfrak{her}_0(N))$ which is also invariant under
the involution $\Lambda \mapsto - \Lambda^\top$.
By choosing an orthogonal basis in $\mathfrak{her}_0(N)$,
this invariance of $\g$ translates into a matrix representation of $\g$
which is stable under $X \mapsto - X^\dagger$. Then Proposition 1.59
in \cite{Knapp02} implies that $\g$ is reductive and thus 
it decomposes into a direct sum of its centre $\mathfrak{z}$ and
its semi-simple commutator ideal $\g_0 := [\g,\g]$,
i.e.~$\g = \mathfrak{z} \oplus \g_0$. Since $\ad{\su(N)}$, is
contained in $\g$, the centre $\mathfrak{z}$ is either trivial or
$\R \cdot \unity$. 
%
%
Thus, similar to Corollary 7.10 in \cite{Knapp02}, one can show
that $\bG := \langle \exp \g \rangle$ is a closed connected subgroup
of $GL(\mathfrak{her}_0(N))$. 
%
%
Therefore, Theorem \ref{thm:kp-Liewedge} applies to $\bG$. In
particular, $\mathfrak k$ and $\mathfrak p$ yield
the required eigenspace decomposition of $\g$. 
Hence we conclude that 
$\exp(- \mathfrak c) \cdot \langle \exp \mathfrak k \rangle =
\exp(- \mathfrak c) \cdot \Ad{SU(N)}$ is a Lie subsemigroup of
$GL(\mathfrak{her}_0(N))$ with Lie wedge
$(-\mathfrak c) \oplus \ad{\su(N)}$.
Thus the result follows. \hfill$\blacksquare$


\medskip

The previous findings suggest the following procedure to
compute or at least to approximate the Lie wedge of
$\overline{\bP}_\Sigma$:

\begin{enumerate}
\item[(i)]
Check, whether $\Gamma_L$ is self-adjoint (implying
positive semidefiniteness for $\Gamma_L$). This is
for example the case, if all $V_k$ in Eqn.~\eqref{eqn:LindKoss1}
are Hermitian or, equivalently, if the Kossakowski-Lindblad term
can be rewritten as a sum of double commutators,
cf.~Eqn.~\eqref{eqn:doublebracket}.
\item[(ii)]
If (i) holds, find the smallest cone $\mathfrak c$ containing
$\Gamma_L$ and satisfying the conditions of Theorem
\ref{thm:opensys-Liewedge}. 
\end{enumerate}

\noindent
Note that the above procedure yields but an {\em outer approximation}
of the Lie wedge. In general, further arguments are necessary
to obtain equality.
For the generic two-level system in \cite{Alt03}, however,
equality can be proven as the following result shows.

\begin{figure*}[Ht!]
\mbox{\sf (a)\hspace{90mm}(b)\hspace{70mm}}\\
\includegraphics[scale=0.4]{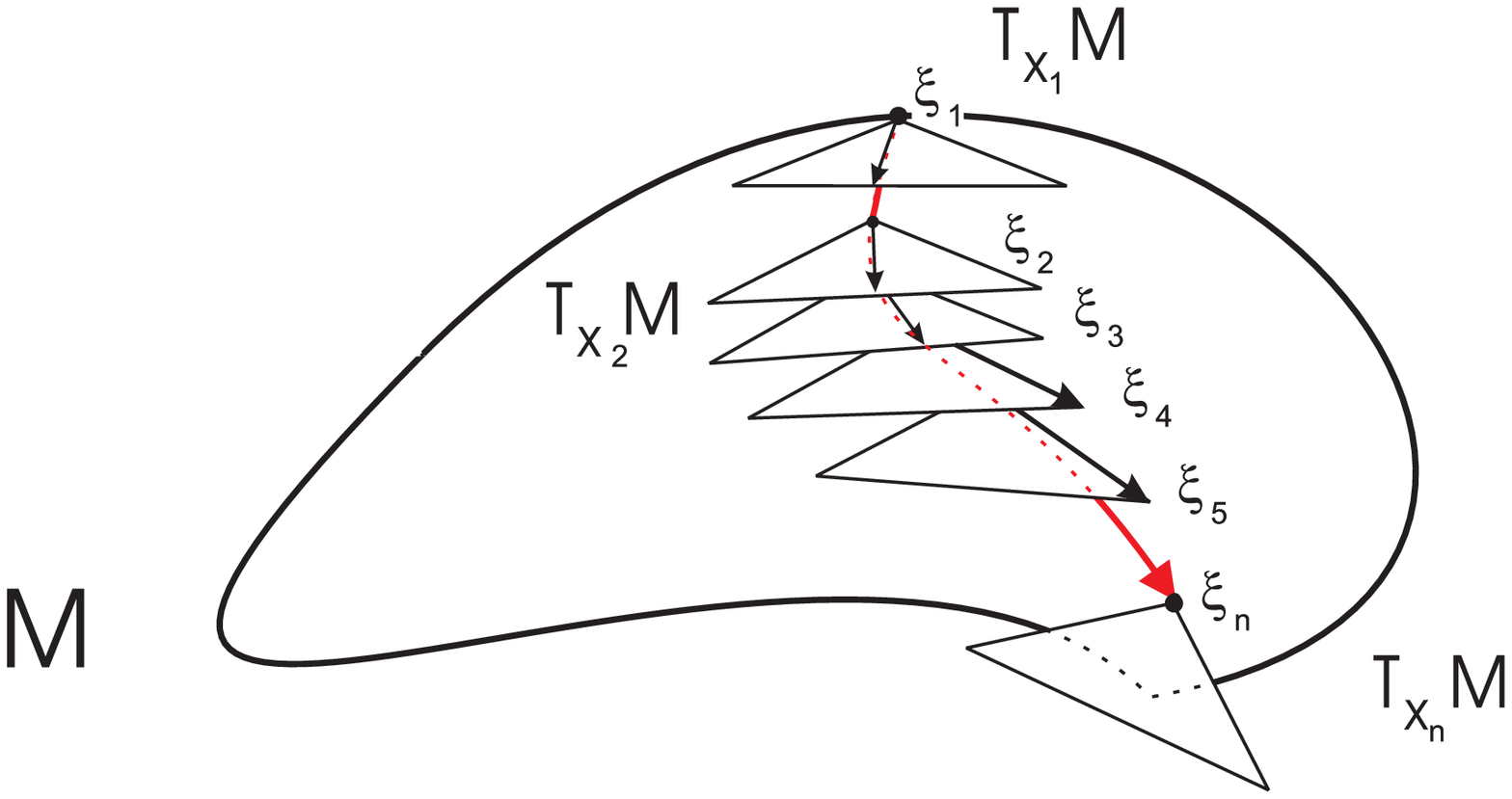}\hspace{22mm}
\raisebox{4mm}{\includegraphics[scale=0.4]{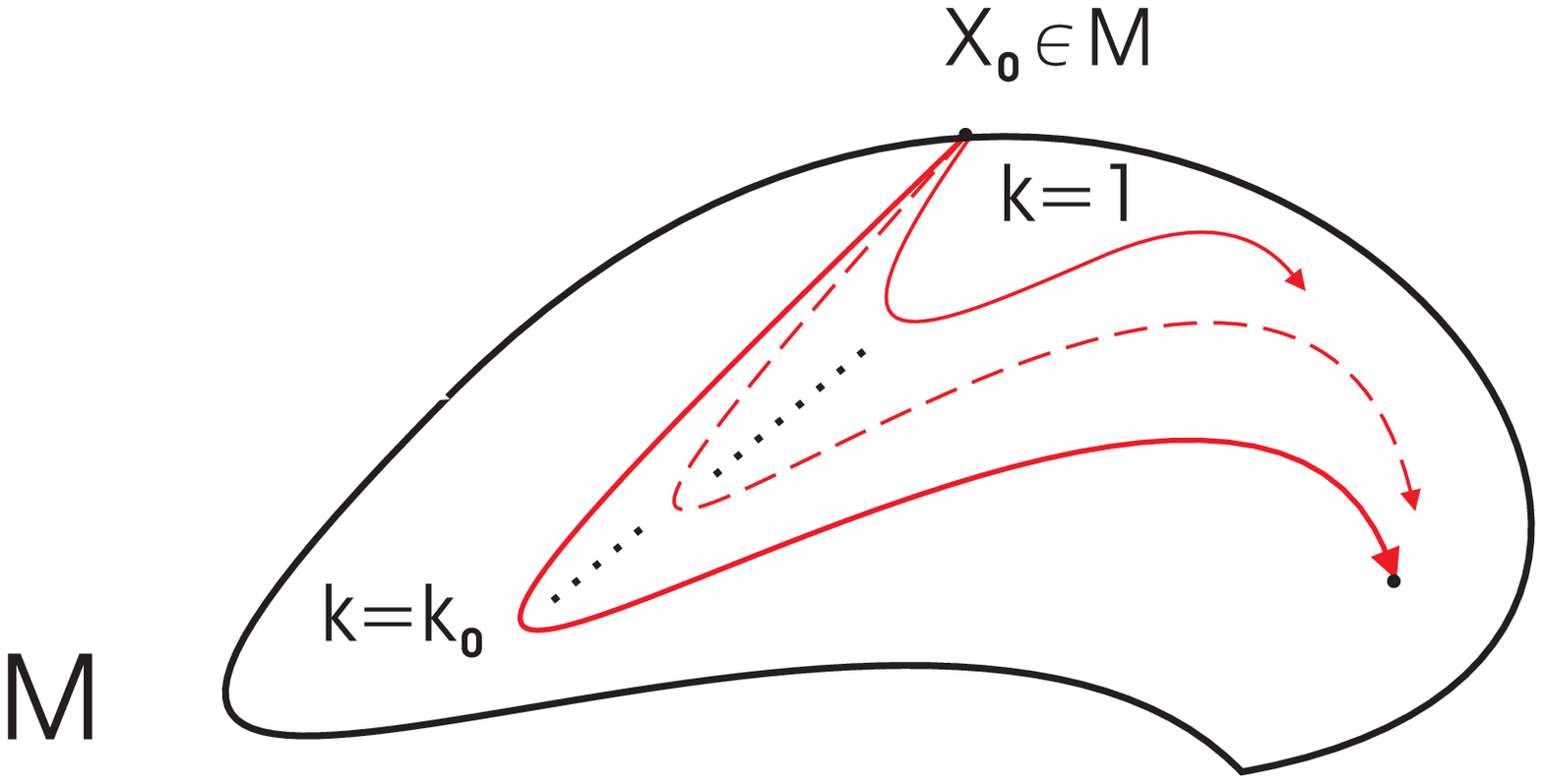}}\\[4mm]
\caption{ \label{fig:mfds_control}(Colour online)
Steering dynamics of open relaxative systems represented by
semigroup actions on a state space manifold $M$:
(a) gradient-like method on the reachable set $\reach(\rho)$
itself; 
admissible directions are confined to directions available
in the Lie wedge;
(b) optimal control approach as an `\/implicit method\/' on the reachable
set $\reach(\rho)$ brought about by a gradient flow on the set of
control amplitudes as in Fig.~3 of Ref.~\cite{SGDH08}.
Note that in (b) the entire trajectory at all points in time is
updated from $k\mapsto k+1$ thus exploring more directions than in
(a), which may be an advantage over local gradient-like methods
in open systems.}
\end{figure*}
\begin{corollary}
\label{cor:opensys-Liewedge}
Let $(\Sigma)$ be a unital {\sc h}-controllable two-level
system with generic Kossakowski-Lindblad term $\Gamma_L$.
Then, the Lie subsemigroup $\overline{\mathbf P}_\Sigma$
coincides with
\begin{equation*}
\overline{\mathbf P}_{\Sigma} = 
\exp(- \mathfrak{c}) \cdot \Ad{SU(2)}
\subset {\mathbf C_0}\big(\mathfrak{her}_0(2)\big)
\end{equation*}
where
$\mathfrak{c}$ denotes the convex cone
\begin{equation}
\label{eqn:cone}
\mathfrak{c} :=
\mathrm{conv}
\big\{ \lambda \Theta \Gamma_L \Theta^\top
\;|\; \lambda \geq 0,\, \Theta \in  \Ad{SU(2)}\big\}
\end{equation}
contained in the set of all positive semidefinite elements
in $\gl(\mathfrak{her}_0(2))$, cf.~Remark \ref{rem:contract}.
Here, $\Theta^\top$ denotes the adjoint operator of $\Theta$ with
respect to the Hilbert-Schmidt inner product on $\mathfrak{her}_0(2)$.
Moreover, the Lie wedge of $\overline{\mathbf P}_{\Sigma}$ is given by
$(-\mathfrak{c}) \oplus \ad{\su(2)}$.
\end{corollary}

{\bf Proof.} {\sc h}-controllability of the system
implies that $\ad{\su(2)}$ is contained in 
$\rL(\overline{\mathbf P}_{\Sigma})$. 
Moreover, for $N=2$ it is known that 
$\Gamma_L\big|_{\mathfrak{her}_0(2)}$ is a positive
semidefinite operator of $\gl(\mathfrak{her}_0(2))$.
Thus Theorem \ref{thm:opensys-Liewedge} applied to the cone
$\mathfrak{c}$ given by Eqn.~\eqref{eqn:cone} yields 
$\overline{\mathbf P}_{\Sigma} \subset 
\exp(- \mathfrak{c}) \cdot \Ad{SU(2)}$.
For the converse inclusion, we refer to a standard convexity result 
on Lie saturated systems, cf.~\cite{Jurdjevic97}.
\hfill$\blacksquare$

\medskip

The geometry of reachability sets under contraction semigroups is
illustrated and summerised in Fig.~\ref{fig:reach_WH_g}.

In general, it is quite intricate to show that outer
approximations of the Lie wedge $\rL(\overline{\mathbf P}_{\Sigma})$
derived from Theorem \ref{thm:opensys-Liewedge} 
in fact coincide with $\rL(\overline{\mathbf P}_{\Sigma})$.
To the best of our knowledge, no efficient procedure
to explicitly determine the global Lie wedge of Eqn.~\eqref{eqn:master2}
does exist. Thus, for optimisation tasks on $\reach(\rho)$,
one currently has to resort to standard optimal control methods.
A straightforward and robust algorithm is mentioned in the
final section. Moreover, a new approach based on an approximation
of $\rL(\overline{\mathbf P}_{\Sigma})$ is sketched.


\section{Relation to Optimisation Tasks}

We follow \cite{SGDH08} in considering optimisation tasks that
come in two scenarios, see also Fig.~\ref{fig:mfds_control}: 
(a) {\em abstract optimisation} over the
reachable set and (b) {\em optimal control} of a dynamic system
specified by its equation of motion (e.g.~of Kossakowski-Lindblad form).
More precisely, an abstract optimisation task means the problem of
finding the global optimum of a given quality function $f$ over the
reachable set of an initial state $\rho$ (independently of the
controls that may drive the system to the desired optimum). In contrast,
a problem is said to be a dynamic optimisation task if one is interested
in an explicit (time dependent) `\/optimal\/' control $u_*$ that steers 
the system as closely as possible to a desired final state, where
`\/optimal\/' can be time- or energy-optimal etc.

In cases where the reachable set $\reach(\rho)$ can be
characterised conveniently---as, for instance, in closed quantum systems
where it is completely characterised by the system
Lie algebra so that $\reach(\rho)$ coincides with the system 
group orbit--- numerical methods from non-linear optimisation (on manifolds)
are appropriate to solve abstract optimisation tasks on $\reach(\rho)$.
Details have been elaborated in \cite{SGDH08}.
However, in open quantum systems a satisfactory characterisation of
the reachable set $\reach(\rho)$---e.g., via Lie algebraic
methods---is currently an unsolved problem. Thus numerical methods
designed for optimal control tasks (b) may serve as handy
substitutes to solve also abstract optimisation tasks (a) on
$\reach(\rho)$.

To be more explicit, we consider the Kossakowski-Lindblad equation
\eqref{Eq:CLKE} with controlled Hamiltonian \eqref{eqn:H_defs}
in superoperator representation. We are faced with a system taking the form
of a standard {\em bilinear control system} ($\Sigma$) for
$\vec\rho \in \C^{N^2}$ reading
\begin{equation}
\label{eqn:BilControl}
\vec \dot \rho = 
\big(A_0 + \sum_{j=1}^m u_j A_j\big) \; \vec\rho
\end{equation}
with 
drift term $A_0 := - \ri (\unity_N \otimes H_d - H^{\top}_d \otimes \unity_N)
- \widehat\Gamma_L$,
control directions
$A_j := - \ri (\unity_N \otimes H_j-H^{\top}_j \otimes \unity_N)$,
and control amplitudes $u_j\in\R{}$, while $\widehat{\Gamma}_L$
is given by Eqn.~\eqref{eqn:LindKoss2}.
Then an optimal control task
boils down to maximising a quality functional 
with respect to some finite dimensional function space, e.g.,
piecewise constant control amplitudes (for details see \cite{SGDH08}
Overview Section).
Clearly, one can reduce the size of system \eqref{eqn:BilControl}
by choosing a coherence-vector representation
instead of a superoperator representation without changing the principle
approach.

In this context, we would like to point out a remarkable
interpretation of $\rL(\overline{\mathbf{P}}_\Sigma)$. The
method just outlined may lead to a (discretised) unconstrained
gradient flow on some high-dimensional $\R^{m}$.
While the `\/local\/' search directions (pulled back to state space)
are confined to directions available in the `\/local\/' Lie wedge of
Eqn.~\eqref{eqn:master2}, i.e. to the smallest Lie wedge generated
by $A_0$ and $u_j A_j$, $u_j \in \R$, the entire method nevertheless
allows to vary the final point $\rho(T)$ within an open
neighbourhood of $\reach(\rho)$, cf. Fig.~\ref{fig:mfds_control}(b).
In contrast, a gradient-like method on the reachable set itself
similar to the one for closed systems, 
but with search directions constrained to the (local) Lie wedge
would in general fail, cf. Fig.~\ref{fig:mfds_control}(a). 

\subsubsection*{Outlook: An Algorithm Exploiting the Lie-Wedge}

Yet, combining both methods yields a new
approach to abstract optimisation tasks: (i) First determine an
\emph{inner} approximation $\mathfrak{c}$ of the Lie wedge.
(ii) Then, choose $n \in \N$ and define a map from the $n$-fold
cartesian product $\mathfrak{c} \times \cdots \times \mathfrak{c}$
to $\R$ by
$(\Omega_1, \dots, \Omega_n) \mapsto f(\e^{\Omega_n}\cdots\e^{\Omega_1})$.
Optimise this function over the {\em convex set}
$\mathfrak{c} \times \cdots \times \mathfrak{c}$
and increase $n$ if necessary. We do expect that the performance 
of such an approach improves the better the approximation of the
Lie wedge is. In particular, the length of the necessary products
$\e^{\Omega_n}\cdots\e^{\Omega_1}$ will significantly decrease if
$\mathfrak{c}$ is a good approximation to $\rL(\overline{\mathbf{P}}_\Sigma)$.
Thus even for numerical aspects knowing the
Lie wedge is of considerable interest. ---
With these remarks we will turn to other points pertinent in practice. 


\subsection*{Practical Implications for Current Numerical Optimal Control}

The above considerations have further implications for numerical
approaches to optimal control of open systems in the sense of the
dynamic task (b) of the previous section. They provide the
framework to understand why time-optimal control makes sense in
certain {\sc wh}-controllable systems, whereas all other situations
ask for explicitly taking the Kossakowski-Lindblad master equation
into account. Consider three scenarios: (i) open quantum systems
that  {\sc wh}-controllable with almost uniform decay rate,
(ii) generic open systems with known Markovian (or non-Markovian)
relaxation characteristics, and
(iii) open systems with unknown relaxation behaviour.

In the simple case (i) of a {\sc wh}-controllable system with
almost uniform decay rate $\gamma$, $\Gamma_L$ approximately acts on
$\mathfrak{her}_0(N)$ as scalar $\gamma\unity$.
Now assume that by numerical optimal control a build-up top curve
$g(T)$ (value function) of maximum obtainable quality against total
duration $T$ was calculated for the corresponding closed system 
with $\Gamma_L = 0$.
Moreover, let $T_*$ denote the smallest time allowing for a quality 
above a given error-correction threshold. 
Together with the uniform decay rate $\gamma$ this
already provides all information if the quality function
depends linearly on $\rho(T)$.
Hence determining $T'_*  := \argmax \{ g(T)\cdot e^{-\gamma T}\}$
gives the optimal time for the desired solution.
More coarsely if $T'_* \simeq T_*$, {\em time-optimal controls
for the closed system} are already a good guess for steering
a {\sc wh}-controllable system with almost uniform decay rate.

For case (ii), when the Kossakowski-Lindblad
operator is known, but generically does not commute with all Hamiltonian
drift and control components, it is currently most advantageous to use
numerical optimal control techniques based on the Master equation with specific
Kossakowski-Lindblad terms as has been illustrated in \cite{PRL_decoh}.
The importance of including the Kossakowski-Lindblad terms roots
in the fact that their non-commutative interplay with the Hamiltonian
part actually introduces new directions in the semigroup dynamics.
Likewise, in \cite{PRL_decoh2}, we treated the optimal control task
of open quantum systems in a non-Markovian case, where a qubit
interacts in a non-Markovian way with a two-level-fluctuator, which
in turn is dissipatively coupled to a bosonic bath in a Markovian way.

Clearly, the case of entirely unknown relaxation characteristics
(iii), where e.g., model building and system identification of the
relaxative part is precluded or too costly, is least expected to
improve by suitable open-loop controls, if at all. 
Yet in \cite{PRL_decoh} we have demonstrated that guesses
of time-optimal control sequences (again obtained from the analogous
closed system) may---by sheer serendipity---be apt to cope with
relaxation. In practice, this comes at the cost of making sure a
sufficiently large family of time-optimal controls is ultimately
tested in the actual experiment for selecting among many
optimal-control based candidates by trial and error. --- Since
this procedure is clearly highly unsatisfactory from a scientific viewpoint, 
efficient methods of determining pertinent decay parameters are highly desirable.


\section*{CONCLUSIONS}

Optimising quality functions for open quantum dynamical processes
as well as determining steerings in concrete experimental settings
that actually achieve these optima is tantamount to exploiting and
manipulating quantum effects in future technology.

To this end, we have recast the structure of completely positive
trace-preserving maps describing the time evolution of open quantum
systems in terms of \emph{Lie semigroups}. 
On an abstract level, the semigroups of completely positive
operators may thus be seen as a special instance within the more general
theory of invariant cones \cite{Vinberg80,HHL89}.
Here, we have
identified the set of Kossakowski-Lindblad generators as
\emph{Lie wedge}: the tangent cone at the unity
of the subsemigroup of all invertible, completely positive, and
trace-preserving operators coincides with the set of
Kossakowski-Lindblad operators.  

In particular, (in the connected component of the unity)
invertible quantum channels are time dependent Markovian, if they belong to 
the {\em Lie semigroup} generated by the {\em Lie wedge} of all Kossakowski-Lindblad
operators.
Moreover, a time dependent Markovian channel specialises to a 
time {\em in}\/dependent Markovian one, if the Lie wedge
of an associated semigroup shows the stronger structure of
a {\em Lie semialgebra}. ---
Likewise, in time dependently controlled open systems the existence of {\em effective Liouvillians} 
that comply with the dynamics given by the Master equation is linked to Lie-semialgebra structures.

In view of controlling open quantum systems, reachable sets have been
described in the same framework. Compared to closed systems,
the structure of  reachable sets of open systems has turned out
to be much more delicate. 
To this end, we have introduced the terms {\em Hamiltonian controllability}
and {\em weak Hamiltonian controllability} replacing the standard notion of controllability,
which fails in open quantum systems whenever the control restricts to
the Hamiltonian part of the system.
For simple cases, we have characterised Hamiltonian controllability
and weak Hamiltonian controllability. These definitions also allow
for characterising the conditons under which time-optimal controls
derived for the associated closed systems already give good
approximations in quantum systems that are actually open.
In the generic case, however, obtaining optimal controls requires numerical
tools from optimal control theory based on the full knowedge of the
system's parameters in terms of its Kossakowski-Lindblad master equation. 

Finally, we have outlined a new algorithmic approach making explicit use
of the Lie wedge of the open system. In cases simple enough to allow
for a good approximation of their respective Lie wedges, a target
quantum map can then be least-squares approximated by a product with comparatively few
factors each taking the form of an exponential of some 
Lie-wedge element.


Since the theory of \emph{Lie semigroups} has only scarcely been used
for studying the dynamics of open quantum systems, the present work
is also meant to structure and trigger further developments.
E.g., the above considerations on $\mathfrak k$-$\mathfrak p$
decompositions may serve as a framework to describe the interplay
of Hamiltonian coherent evolution
and relaxative evolution: this interplay gives
rise to new coherent effects. Some of them relate to 
well-established observations like, e.g., the Lamb-shift \cite{Lamb47} or
dynamic frequency shifts in magnetic resonance \cite{Abragam, Werbelow78, Bruschweiler96},
while others form the basis to very recent findings such as dephasing-assisted quantum transport
in light-harvesting molecules \cite{Mohs08,Reb08qt1,Reb08qt2,Plenio08,Plenio09}. 

\medskip


\medskip
\begin{acknowledgments}
This work was supported in part by the integrated EU programme QAP
and by {\em Deutsche Forschungsgemeinschaft} (DFG) in the
collaborative research centre SFB 631. 
We also gratefully acknowledge
support and collaboration enabled within the two International 
Doctorate Programs of Excellence
{\em Quantum Computing, Control, and Communication} (QCCC) as well as
{\em Identification, Optimisation and Control with Applications in
Modern Technologies} by the Bavarian excellence network ENB.
We wish to thank Prof.~Michael Wolf 
for clarifying discussions
on non-Markovian quantum channels \cite{Wolf08pc}, while
Prof.~Bernard Bonnard (Universit{\'e} de Bourgogne) pointed out
some useful older literature.
T.S.H. is grateful to Prof.~Hans Primas (ETH-Zurich) for his early
attracting attention to completely positive semigroups
and for valuable exchange. 
\end{acknowledgments}

\bibliography{control21}
\end{document}